\documentclass[a4paper,11pt]{article}
\pdfoutput=1 

\usepackage{jheppub} 

\usepackage[T1]{fontenc} 
\usepackage{comment}

\title{Chiral higher-spin double copy}

\author{Dmitry Ponomarev}

\affiliation{Institute for Theoretical and Mathematical Physics,\\
Lomonosov Moscow State University, Lomonosovsky avenue, Moscow, 119991, Russia}
\affiliation{I.E. Tamm Theory Department, Lebedev Physical Institute,\\
 Leninsky avenue, Moscow, 119991, Russia}

\emailAdd{ponomarev@lpi.ru}

\abstract{We construct the double copy of the chiral higher-spin theory. It is a Lorentz invariant theory with the little group spectrum given by the tensor square of  the chiral higher-spin theory spectrum. Moreover, its interactions factorise in close analogy with the way interactions factorise  in lower-spin double-copy theories. We also propose theories, which can be viewed as products of self-dual Yang-Mills theory, self-dual gravity and chiral higher-spin theories taken in different combinations and powers.}

\begin{document} 
\maketitle
\flushbottom

\section{Introduction}

Relations between different theories have always played a central role in theoretical physics. 
One of the most intriguing relations of this type is color-kinematics duality. Taking its origin in string theory \cite{Kawai:1985xq} it was then extended to field theory, in which it famously relates gauge theory and gravity \cite{Bern:2008qj,Bern:2010ue,Bern:2010yg}. In the latter case color-kinematics duality states that amplitudes in Yang-Mills theory once properly rearranged  give rise to gravitational amplitudes via a certain squaring procedure referred to as the double copy. Besides that, color-kinematics duality hints towards the existence of the so-called kinematic symmetry, which controls gauge theory amplitudes in a way closely analogous to the way these are controlled by the color symmetry. The associated kinematic constraints are known as the BCJ relations. In recent years color-kinematics duality received considerable attention, which resulted in various generalisations and extensions. In particular, it applies to a wide range of theories, connecting them into an intricate duality web, see \cite{Bern:2019prr,Kosower:2022yvp,Adamo:2022dcm} for recent reviews.

Despite the remarkable success of color-kinematics duality,  
some of its underlying structures  remain rather obscure. In particular, the kinematic algebra of Yang-Mills theory is not known.
Similarly, the double copy only works at the level of amplitudes, which, moreover, need to be properly  preprocessed before the procedure can be applied. The situation is dramatically different for self-dual theories due to the fact that both self-dual Yang-Mills theory and self-dual gravity can be reformulated in a way that only cubic vertices are present. 
In this form the self-dual Yang-Mills vertex explicitly factorises into the product of the color and the kinematic structure constants. This feature allows one to identify the self-dual Yang-Mills kinematic algebra as the algebra of area-preserving diffeomorphisms \cite{Monteiro:2011pc}. Besides that, the self-dual gravity vertex manifestly has the form of the kinematic part of the Yang-Mills vertex squared.
Thus, both constituents of color-kinematics duality -- the double copy and the BCJ relations -- become manifest in the self-dual sector even at the level of Lagrangian. For further developments within this framework, see e.g. \cite{Boels:2013bi,Armstrong-Williams:2022apo,Monteiro:2022nqt,Doran:2023cmj,Ivanovskiy:2024ads}.

Another special feature of the self-dual setup is that it allows one to construct higher-spin theories at the same time avoiding the necessity to deal with non-localities. 
Based on the earlier analysis of \cite{Metsaev:1991mt,Metsaev:1991nb} such manifestly local chiral higher-spin theories were suggested in \cite{Ponomarev:2016lrm}. A related result was obtained in \cite{Devchand:1996gv}, where it was shown that self-dual Yang-Mills theory admits supersymmetrisation with an arbitrary number of supersymmetries, which, obviously, requires fields of arbitrarily high spin to be present in the spectrum.
Since chiral higher-spin theories were suggested, they attracted considerable interest. In particular, quantum corrections in these theories were computed \cite{Skvortsov:2018jea,Skvortsov:2020wtf}, chiral higher-spin theories were covariantised and deformed to anti-de Sitter space \cite{Krasnov:2021nsq,Sharapov:2022faa,Sharapov:2022wpz}, their connection to twistors was explored \cite{Herfray:2022prf,Adamo:2022lah,Neiman:2024vit}, flat-space holographically dual pair with the chiral higher-spin theory as the bulk theory was found \cite{Ponomarev:2022ryp,Ponomarev:2022qkx} and extensions to higher dimensions were suggested \cite{Basile:2024raj}.

Quite remarkably,  chiral higher-spin theories display some form of the BCJ structure \cite{Ponomarev:2017nrr,Monteiro:2022lwm,Monteiro:2022xwq}. More precisely, they  feature only a single cubic  vertex, which factorises into the product of the kinematic part of the Yang-Mills vertex and the structure constant of the chiral higher-spin algebra\footnote{Unlike \cite{Monteiro:2022lwm,Monteiro:2022xwq} we prefer to regard this structure as a version of the BCJ relations, not of the double copy. By the generalised double copy we will understand a different pattern, which will be explained below.}. By generalising the lower-spin construction of \cite{Park:1990fp}  it was  shown in  \cite{Ponomarev:2017nrr} that this structure implies that any theory possessing it can be rephrased as a self-dual Yang-Mills theory and as a 2d principal chiral model with, possibly, space-time symmetry in place of internal symmetry. 
This, in turn, implies that theories of this type are integrable and, as a result, the associated tree level amplitudes vanish. The same conclusions were reached differently in \cite{Chacon:2020fmr,Monteiro:2022lwm,Monteiro:2022xwq}, where this discussion was also connected to the Ward conjecture
\cite{Ward:1985gz}.
Besides that, in \cite{Ponomarev:2017nrr}  it was also argued that the aforementioned  structure 
is a universal feature of chiral theories and follows from Lorentz invariance alone.

Considering that  chiral higher-spin theories fit seamlessly into the BCJ pattern in the form explained above, it is natural to 
wonder whether some extension of the double copy -- the second ingredient of color-kinematics duality -- applies to these theories as well. 

In the present paper we address the problem of the chiral higher-spin double copy in the following way. To start, we consider a theory with the spectrum, which at the little group level is given by the tensor product of the spectra of two chiral higher-spin theories. Putting differently, the fields in the double-copy theory are labelled by two integers, each associated with a helicity of the single copy theory. Then the sum of these labels is field's helicity in the double-copy theory, while their difference is an additional parameter labelling spectrum degeneration. As the next step, we make this theory interacting by adding chiral cubic vertices. To mimic the  double copy procedure for lower-spin self-dual theories we require that these vertices factorise into two factors each depending only on one set of helicities.
 Finally, we use the light-cone deformation procedure to impose Lorentz invariance and, thus, completely fix the double-copy chiral higher-spin theory\footnote{Let us highlight that Lorentz invariance is the key constraint that we impose when constructing the double-copy theory. This is another difference between our approach and that of \cite{Chacon:2020fmr,Monteiro:2022lwm,Monteiro:2022xwq}, which together with Lorentz invariant theories also considered theories for which Lorentz symmetry is  violated.}. We also extend this procedure to higher powers of chiral higher-spin theories, as well as to their products with powers of the self-dual Yang-Mills and self-dual gravity factors. 
 
We would like to emphasise that  chiral higher-spin theories that we will construct following the procedure outlined above are new. Unlike the original chiral higher-spin theory, which features a field of each helicity only once, its  copies have degenerate spectra. In particular, the double copy of the chiral higher-spin theory involves fields of each spin infinitely many times, which makes this theory reminiscent of string theory. Potential connections to string theory and the utility of such spectra in the context of higher-spin symmetry breaking will be further discussed in the concluding section.

This paper is organised as follows. In section \ref{sec:2} we review color-kinematics duality for self-dual theories as well as motivate our setup for the chiral higher-spin double copy. Next, in section \ref{sec:3} we briefly review the necessary background on the light-cone deformation procedure. 
Then, in section \ref{sec:4}, focusing on theories with no internal symmetry, we construct the double copy of the chiral higher-spin theory as well as its multiple copies and products with self-dual Yang-Mills theory and self-dual gravity. Fields in the adjoint representation of some internal Lie algebra are treated in section \ref{sec:5} in a similar way. In section \ref{sec:syma} we discuss how the double copy works at the level of symmetry algebras. Finally, we conclude in section \ref{sec:6}. Appendices contain our conventions as well as some technical results.

\section{Self-dual color-kinematics duality}
\label{sec:2}

In the present section we  briefly review how color-kinematics duality works for self-dual theories in known cases. We will make emphasis on some of its structural features which will be kept in the following sections in which the chiral higher-spin double copy and other its extensions will be considered.

\subsection{The original  pattern}

The action for self-dual Yang-Mills theory in the light-cone gauge can be formulated in terms of helicity $+1$ and helicity $-1$ fields, $\Phi^{+1}$ and $\Phi^{-1}$, as follows \cite{Cangemi:1996rx,Chalmers:1996rq,Bardeen:1995gk}
\begin{equation}
\label{3aug1}
S_{SDYM}= \int d^4 x (\Phi^{a|-1}\Box \Phi^{a|+1} + gf^{abc}\frac{\bar{\mathbb{P}}\beta_1}{\beta_2\beta_3} \Phi^{a|-1}\Phi^{b|+1}\Phi^{c|+1}).
\end{equation}
Here  $f^{abc}$ are the internal Lie algebra structure constants,
 \begin{equation}
 \label{3aug2}
 \begin{split}
 \bar{\mathbb{P}} \equiv \bar\partial_1\beta_2-\bar\partial_2 \beta_1, \qquad  \beta_i = \partial^+_i
  \end{split}
 \end{equation}
and lower indices of derivatives refer to the field number the derivative acts upon: for example, $\beta_1$ acts on the first field and $\bar\partial_2$ acts
on the second one\footnote{Here we use the standard conventions employed in the literature on the light-cone deformation procedure in the higher-spin context. These can be easily connected to other conventions, e. g. those of \cite{Monteiro:2011pc}. Details on our conventions and on their connection to the conventions of \cite{Monteiro:2011pc} can be found in appendix \ref{app:0}.}.

The action for self-dual gravity can be written in a similar way \cite{Plebanski:1975wn}
\begin{equation}
\label{3aug5}
S_{SDGR}= \int d^4 x (\Phi^{-2}\Box \Phi^{+2} + \kappa \frac{\bar{\mathbb{P}}^2\beta^2_1}{\beta^2_2\beta^2_3} \Phi^{-2}\Phi^{+2}\Phi^{+2}),
\end{equation}
where $\Phi^{+2}$ and $\Phi^{-2}$ are fields of helicities $+2$ and $-2$ respectively.

Color-kinematics duality for actions (\ref{3aug1}) and (\ref{3aug5}) manifests itself in the following way. Firstly, the vertex in the Yang-Mills case factorises into 
\begin{equation}
\label{3aug6}
c \equiv f^{abc} \qquad \text{and} \qquad n\equiv \frac{\bar{\mathbb{P}}\beta_1}{\beta_2\beta_3},
\end{equation}
where $n$ is the color Lie algebra structure constant, while $n$ defines the kinematic algebra of area-preserving diffeomorphisms \cite{Monteiro:2011pc}. 
For the demonstration of the latter fact in our notations, see \cite{Ponomarev:2017nrr}. Secondly, the vertex for self-dual gravity can be obtained from that of self-dual Yang-Mills by the replacement $c\to n$, which is a manifestation of its double-copy structure.

Typically, to further highlight the double-copy pattern one adds the zeroth copy, which corresponds to the bi-adjoint scalar theory
\begin{equation}
\label{3aug7}
S_{bi-sc}= \int d^4 x (\frac{1}{2}\Phi^{a\tilde a| 0}\Box \Phi^{a\tilde a|0} + gf^{abc}\tilde f^{\tilde a\tilde b\tilde c}\Phi^{a\tilde a|0}\Phi^{b \tilde b|0}\Phi^{c\tilde c|0}).
\end{equation}
Here $f$ and $\tilde f$ are the structure constants of two, possibly, different internal Lie algebras. Then, the double-copy procedure connects vertices in three theories
\begin{equation}
\label{3aug8}
\begin{split}
\text{zeroth copy:}& \qquad V_3 = c\tilde c,\\
\text{single copy:}& \qquad V_3 = cn,\\
\text{double copy:}& \qquad V_3 = n n, 
\end{split}
\end{equation}
which can be obtained one from another by replacements of the color and kinematic structure constants. Various generalisations of this pattern were suggested in \cite{Chacon:2020fmr}.

\subsection{Extension with the Kalb-Ramond field and the dilaton}
\label{tpdc}

The double copy is often defined in a slightly different way, which we find more natural and suitable for further generalisations. Namely, instead of regarding pure gravity as the double copy of Yang-Mills theory,  one can consider a double copy which results in gravity  interacting with the dilaton and the Kalb-Ramond field.

This version of the double copy is motivated in the following way. At the little group level the Yang-Mills field transforms in the vector representation. The tensor product of two vector representations results in the direct sum of the symmetric traceless rank-2 tensor, the antisymmetric rank-2 tensor and the scalar. These are precisely the little group representations carried by graviton, the Kalb-Ramond field and the dilaton respectively. Thus, at the level of the spectrum, this version of  the double copy amounts to the tensor square of the little group representation carried by the single-copy theory. 

This discussion can be naturally rephrased in terms of helicities.
The single copy fields can be labelled by a single label, their helicity. For example, for the Yang-Mills field we have
\begin{equation}
\label{3aug9}
\text{single copy:} \qquad \Phi^{+1} \oplus \Phi^{-1}.
\end{equation}
For its double copy, to manifest the tensor product structure, we will use pairs of labels to label fields. For example, for the double copy of Yang-Mills theory one has
\begin{equation}
\label{3aug10}
\text{double copy:} \qquad \Phi^{+1,+1} \oplus \Phi^{+1,-1} \oplus \Phi^{-1,+1} \oplus  \Phi^{-1,-1}.
\end{equation}
Obviously, the sum of labels gives the helicity associated with a given field. If necessary, (\ref{3aug10}) can be, of course, decomposed into combinations associated with graviton, the Kalb-Ramond field and the dilaton. For example, gravity is described by fields $\Phi^{+1,+1}$ and $\Phi^{-1,-1}$. We, however, prefer to keep our fields in the form (\ref{3aug10}), which helps us making the tensor product structure of the spectrum manifest.

For interactions in the self-dual sector one then proceeds as follows. First, we rewrite the action for self-dual Yang-Mills theory  (\ref{3aug1}) in the Bose symmetric way
\begin{equation}
\label{3aug11}
\begin{split}
&S_{SDYM}= \int d^4 x \big(\Phi^{a|-1}\Box \Phi^{a|+1} \\
&\qquad + \frac{g}{3}f^{abc}\sum_{\lambda=\pm 1}\frac{\bar{\mathbb{P}}}{\beta^{\lambda_1}_1\beta^{\lambda_2}_2\beta^{\lambda_3}_3}
(\delta^{\lambda_1}_1\delta^{\lambda_2}_1\delta^{\lambda_3}_{-1}+\delta^{\lambda_1}_1\delta^{\lambda_2}_{-1}\delta^{\lambda_3}_{1}+\delta^{\lambda_1}_{-1}\delta^{\lambda_2}_1\delta^{\lambda_3}_{1})
 \Phi^{a|\lambda_1}\Phi^{b|\lambda_2}\Phi^{c|\lambda_3}\big).
 \end{split}
\end{equation}
Then the double copy gives 
\begin{equation}
\label{3aug12}
\begin{split}
&S_{SDGR+KR+D}= \int d^4 x \big(\Phi^{-1,-1}\Box \Phi^{+1,+1}+\Phi^{-1,+1}\Box \Phi^{+1,-1} \\
&\qquad + \frac{\kappa}{9}\sum_{\lambda,\mu=\pm 1}\frac{\bar{\mathbb{P}}^2}{\beta^{\lambda_1+\mu_1}_1\beta^{\lambda_2+\mu_2}_2\beta^{\lambda_3+\mu_3}_3}
(\delta^{\lambda_1}_1\delta^{\lambda_2}_1\delta^{\lambda_3}_{-1}+\delta^{\lambda_1}_1\delta^{\lambda_2}_{-1}\delta^{\lambda_3}_{1}+\delta^{\lambda_1}_{-1}\delta^{\lambda_2}_1\delta^{\lambda_3}_{1})\\
&\qquad\qquad  \qquad\qquad\qquad  (\delta^{\mu_1}_1\delta^{\mu_2}_1\delta^{\mu_3}_{-1}+\delta^{\mu_1}_1\delta^{\mu_2}_{-1}\delta^{\mu_3}_{1}+\delta^{\mu_1}_{-1}\delta^{\mu_2}_1\delta^{\mu_3}_{1})
 \Phi^{\lambda_1,\mu_1}\Phi^{\lambda_2,\mu_2}\Phi^{\lambda_3,\mu_3}\big),
 \end{split}
\end{equation}
which describes a self-dual theory of gravity interacting with the Kalb-Ramond field and the dilaton. The pattern (\ref{3aug8}) still holds, except that we need to define
\begin{equation}
\label{3aug13}
n'\equiv \sum_{\lambda=\pm 1}\frac{\bar{\mathbb{P}}}{\beta^{\lambda_1}_1\beta^{\lambda_2}_2\beta^{\lambda_3}_3}
(\delta^{\lambda_1}_1\delta^{\lambda_2}_1\delta^{\lambda_3}_{-1}+\delta^{\lambda_1}_1\delta^{\lambda_2}_{-1}\delta^{\lambda_3}_{1}+\delta^{\lambda_1}_{-1}\delta^{\lambda_2}_1\delta^{\lambda_3}_{1}).
\end{equation}

If necessary, one can consistently set $\Phi^{1,-1}$ and $\Phi^{-1,1}$ to zero in (\ref{3aug12}), which will lead us back to (\ref{3aug5}) with $\Phi^2=\Phi^{1,1}$ and $\Phi^{-2}=\Phi^{-1,-1}$. We, however, prefer to regard the complete theory  (\ref{3aug12}) as the double copy of self-dual Yang-Mills theory, as this version of the double copy seems more natural and more suitable for generalisations.

\subsection{Alternative version of the BCJ relations}
\label{BCJsd}

As was mentioned in the introduction, chiral higher-spin theories also fall into the BCJ relations pattern, which is, however, somewhat different from the one we discussed above. We will briefly review it below.

The action of the chiral higher-spin theory in the colorless case reads as follows
\begin{equation}
\label{4aug1}
S_{CHS}= \frac{1}{2}\sum_{\lambda \in \mathbb{Z}} \int d^4x \Phi^{\lambda}\Box \Phi^{-\lambda} + g\sum_{\lambda_i \in \mathbb{Z}}\frac{l^{\lambda_1+\lambda_2+\lambda_3-1}}{(\lambda_1+\lambda_2+\lambda_3-1)!}\int d^4 x \frac{\bar{\mathbb{P}}^{\lambda_1+\lambda_2+\lambda_3}}{\beta_1^{\lambda_1}\beta_2^{\lambda_2}\beta_3^{\lambda_3}} \Phi^{\lambda_1}\Phi^{\lambda_2}\Phi^{\lambda_3}.
\end{equation}
By appropriately rescaling the coupling constant and sending $l$ to zero, we arrive at
\begin{equation}
\label{4aug2}
S_{PCHS}= \frac{1}{2}\sum_{\lambda \in \mathbb{Z}} \int d^4x \Phi^{\lambda}\Box \Phi^{-\lambda} + g\sum_{\lambda_i \in \mathbb{Z}}\delta^{\lambda_1+\lambda_2+\lambda_3}_2\int d^4 x \frac{\bar{\mathbb{P}}^{2}}{\beta_1^{\lambda_1}\beta_2^{\lambda_2}\beta_3^{\lambda_3}} \Phi^{\lambda_1}\Phi^{\lambda_2}\Phi^{\lambda_3}.
\end{equation}
This theory was referred to as the Poisson chiral higher-spin theory in \cite{Ponomarev:2017nrr}. A closely related theory is also known as higher-spin self-dual gravity \cite{Krasnov:2021nsq}.
Besides that, one can also consider chiral higher-spin theories with fields transforming in the adjoint representations of some internal symmetry algebra. The associated action reads 
\begin{equation}
\label{4aug3}
S_{CCHS}= \frac{1}{2}\sum_{\lambda \in \mathbb{Z}} \int d^4x \Phi^{a|\lambda}\Box \Phi^{a|-\lambda} + gf^{abc}\sum_{\lambda_i \in \mathbb{Z}}\delta^{\lambda_1+\lambda_2+\lambda_3}_1\int d^4 x \frac{\bar{\mathbb{P}}}{\beta_1^{\lambda_1}\beta_2^{\lambda_2}\beta_3^{\lambda_3}} \Phi^{a|\lambda_1}\Phi^{b|\lambda_2}\Phi^{c|\lambda_3}.
\end{equation}
A closely related theory is also known as higher-spin self-dual Yang-Mills theory  \cite{Krasnov:2021nsq}. Due to a peculiar form of interactions, theories (\ref{4aug2}) and (\ref{4aug3}) admit consistent truncations to  smaller spectra. We will not discuss these truncations below. Additional opportunities arise if one considers matrix-valued fields \cite{Skvortsov:2018jea,Skvortsov:2020wtf}. This option will not be considered here neither. For a nice summary on the list of chiral higher-spin theories, see \cite{Monteiro:2022xwq}.

Theories (\ref{4aug1})-(\ref{4aug3}) manifest the following pattern \cite{Ponomarev:2017nrr,Monteiro:2022lwm,Monteiro:2022xwq}. Firstly, one can notice that 
\begin{equation}
\label{4aug4}
\begin{split}
n_{CHS}&\equiv \frac{l^{\lambda_1+\lambda_2+\lambda_3-1}}{(\lambda_1+\lambda_2+\lambda_3-1)!}\frac{\bar{\mathbb{P}}^{\lambda_1+\lambda_2+\lambda_3-1}}{\beta_1^{\lambda_1-1}\beta_2^{\lambda_2-1}\beta_3^{\lambda_3+1}},\\
n_{PCHS}&\equiv \delta^{\lambda_1+\lambda_2+\lambda_3}_2 
\frac{\bar{\mathbb{P}}}{\beta_1^{\lambda_1-1}\beta_2^{\lambda_2-1}\beta_3^{\lambda_3+1}},\\
n_{CCHS}&\equiv f^{abc}\delta^{\lambda_1+\lambda_2+\lambda_3}_1 
\frac{1}{\beta_1^{\lambda_1-1}\beta_2^{\lambda_2-1}\beta_3^{\lambda_3+1}}
\end{split}
\end{equation}
satisfy the Jacobi identity, thus, defining the associated Lie algebras. Secondly, vertices in theories (\ref{4aug1})-(\ref{4aug3}) can be presented as
\begin{equation}
\label{4aug5}
\begin{split}
\text{CHS:} \qquad V_3 &= n_{CHS} n,\\
\text{PHS:} \qquad V_3 & = n_{PHS}n,\\
\text{CCHS:} \qquad V_3 & = n_{CCHS} n. 
\end{split}
\end{equation}
In other words, vertices in the aforementioned chiral higher-spin theories factorise into the kinematic part of the self-dual Yang-Mills vertex and a structure constant of a suitable Lie algebra. 

Let us emphasise that self-dual Yang-Mills theory and self-dual gravity also fall into pattern (\ref{4aug5}). In \cite{Ponomarev:2017nrr} it was shown that this structure of vertices is a consequence of Lorentz invariance. We will further discuss this issue below.
 Besides that, it was shown, that this structure implies that all theories featuring it can be rewritten as self-dual Yang-Mills theories with a, possibly, space-time symmetry algebra as gauge algebra\footnote{The idea that gauge theories should be regarded as versions of Yang-Mills theory is an important part of Cartan's approach to gravity \cite{Ortin:2015hya} and of its higher-spin generalisation  \cite{Vasiliev:1980as,Aragone:1980rk}.}. Alternatively, all these theories can be rewritten in the form of a principal chiral model, in which the gauge algebra can be made to act only in the internal space. Based on that one can argue that all these theories are integrable. 
 Structure (\ref{4aug5}), as well as its connection to integrability was also highlighted in \cite{Chacon:2020fmr,Monteiro:2022lwm,Monteiro:2022xwq} from a slightly different perspective.

The example of the bi-adjoint scalar does not fit into the pattern (\ref{4aug5}). At the technical level this happens due to this example being too degenerate: Lorentz invariance within the light-cone deformation procedure can be achieved even without the necessity to impose (\ref{4aug5}). This will be illustrated below. Let us also note, that the bi-adjoint scalar  theory is not integrable.

The structure (\ref{4aug5}) was referred to as the double copy in \cite{Monteiro:2022lwm,Monteiro:2022xwq}. Despite this is mainly a terminological issue, we still prefer to regard it as an alternative form of the BCJ relations. Indeed, the original BCJ relations require that the kinematic factor $n$  of the Yang-Mills vertex $f\cdot n$ satisfies the Jacobi identity. In turn, pattern (\ref{4aug5}) applied to gravity implies that the gravitational vertex, $n\cdot n$, divided by the kinematics part of the Yang-Mills vertex, $n$, satisfies the Jacobi identity. Clearly, these two requirements are equivalent, once the double copy is assumed to work, which is why we prefer to regard (\ref{4aug5}) as a version of the BCJ relations.

Having established that the BCJ relations hold for chiral higher-spin theories, we may wonder whether the second part of  color-kinematics duality -- the double copy -- applies to chiral higher-spin theories as well. Unlike \cite{Chacon:2020fmr,Monteiro:2022lwm,Monteiro:2022xwq} we will use a stronger notion of the double copy: besides requiring factorisation of vertices, we will also insist on Lorentz invariance and on factorisation of the spectrum in complete analogy with the lower-spin discussion given in section \ref{tpdc}. 

It is easy to see that the spectrum of chiral higher-spin theories cannot be factorised. Thus, in the double-copy context chiral higher-spin theories can only be  regarded as single copies, while the associated double copies are yet to be constructed. This is the subject of the following sections.

\subsection{Our strategy}
\label{strategy}

Instead of studying the problem of the chiral higher-spin theory double copy alone, we will address the question more generally by considering multiple copies of chiral higher-spin theories as well as their products with self-dual Yang-Mills and self-dual gravity factors. Discussions presented in the previous sections motivate us to suggest the following setup. 

Let $\nu \in {\cal N}$, $\mu \in {\cal M}$, $\rho \in {\cal R}$, $\dots$ be the spectra of helicities of individual theories. 
We will consider theories with the little group spectra given by the tensor products of the spectra of individual theories. 
These spectra consist of the states
\begin{equation}
\label{16jul12}
(\nu,\mu,\rho,\dots), \qquad \nu \in {\cal N}, \quad \mu \in {\cal M}, \quad \rho \in {\cal R}, \quad \dots
\end{equation}
with the total helicity being
\begin{equation}
\label{16jul13}
h=\nu+\mu+\rho+\dots.
\end{equation}

As for interactions, we will consider chiral theories with cubic vertices only. The lower-spin double copy suggests that these have the factorised form
\begin{equation}
\label{4aug6}
V^{{\cal N}, {\cal M}, {\cal R},\dots}_3 = \tilde{V}_3^{\cal N}\tilde{V}_3^{\cal M}\tilde{V}_3^{\cal R}  \dots.
\end{equation}
Here we supplemented vertices on the right-hand side  with tildes to indicate that these may differ from vertices of individual theories entering the product. In other words, (\ref{4aug6}) only implies that the dependence of the total vertex on $\lambda_i$, $\mu_i$, $\nu_i$, $\dots$ factorises.
Finally, we will require that the resulting theories are Lorentz invariant using the light-cone deformation procedure. 

The requirement of Lorentz invariance in the higher-spin case is known to be very constraining. As we will see below, for all theories we will consider it will fix the cubic vertex (\ref{4aug6}) almost uniquely. This explains why we allowed for the possibility of vertices on the right-hand side of (\ref{4aug6}) to be deformed: otherwise, we would have not found any solutions. Still, as we will see below, these deformations are very minor, thus, the interpretation of the resulting theory as the product of individual ones remains well justified. Let us also remark, that the lower-spin double copy alone does not quite give a natural candidate for the double copy of chiral higher-spin theory or its Poisson limit, thus, some freedom in the ansatz on the right-hand side of (\ref{4aug6}) is inevitable. Indeed, the lower-spin double copy requires to remove the color factor from the Yang-Mills vertex as the first step, while this factor is not even present for chiral higher-spin theories without internal symmetries. This factor is only present for the higher-spin theory (\ref{4aug3}). As we will find below, in this latter case our procedure will result in the double copy theory, which does not require any deformations of vertices or, in other words, our result agrees identically with the expectations from the lower-spin double copy procedure.

\section{Light-cone deformation procedure}
\label{sec:3}

In this section we briefly review the light-cone deformation procedure.

The light-cone deformation procedure is a rather technical subject.  Fortunately, most of the steps that we need to carry out do not rely on the spectrum, so we can use the existing results from the literature. The only analysis that we will have to do separately is solving the consistency conditions appearing at the order $g^2$ in the coupling constant. Below we will give a very brief review of the light-cone deformation procedure, focusing only on the summary of results, which will be relevant for our analysis. An interested reader can find a more comprehensive review in \cite{Ponomarev:2022vjb}.

\subsection{First order consistency conditions}
 
 In the light-cone deformation procedure a field of helicity $\lambda$ is represented by a single-component field $\Phi^\lambda$. The free light-cone action involves fields of opposite helicities and reads
 \begin{equation}
 \label{17jul10}
 S_2= \int d^4 x  \Phi^\lambda \Box \Phi^{-\lambda}.
 \end{equation}
 Despite $\Phi^\lambda$ may seem reminiscent of scalar fields,  among them  only $\Phi^0$ transforms as a scalar, while all other fields have more non-trivial transformations under the action of the Lorentz group. This is, of course, necessary to ensure that $\Phi^\lambda$  describe representations of the appropriate helicities on-shell. With the explicit transformation properties of $\Phi^\lambda$ given one can check that the action (\ref{17jul10}) is Poincare invariant despite this is not manifest.

 The key consistency condition that one needs to impose when adding interactions in the light-cone gauge approach is that the Poincare invariance is not broken. To do that one explicitly constructs  phase space Noether charges associated with each of the symmetry generators and demonstrates that these commute as the respective symmetries do. These constraints are implemented perturbatively, which allows one to construct interacting theories order by order starting from a free theory. 
 
 As a result of the leading order analysis one finds the following cubic vertices \cite{Bengtsson:1983pd,Bengtsson:1986kh,Metsaev:1991mt}
 \begin{equation}
 \label{17jul12}
 \begin{split}
 S_3 &= C^{\lambda_1;\lambda_2;\lambda_3}\int d^4 x \frac{\bar{\mathbb{P}}^{\lambda_1+\lambda_2+\lambda_3}}{\beta_1^{\lambda_1}\beta_2^{\lambda_2}\beta_3^{\lambda_3}} \Phi^{\lambda_1}\Phi^{\lambda_2}\Phi^{\lambda_3}, \qquad \lambda_1+\lambda_2+\lambda_3 >0,\\
  S_3 &= C^{\lambda_1;\lambda_2;\lambda_3}\int d^4 x \frac{{\mathbb{P}}^{-\lambda_1-\lambda_2-\lambda_3}}{\beta_1^{-\lambda_1}\beta_2^{-\lambda_2}\beta_3^{-\lambda_3}} \Phi^{\lambda_1}\Phi^{\lambda_2}\Phi^{\lambda_3}, \qquad \lambda_1+\lambda_2+\lambda_3 <0,\\
   S_3 &= C^{0;0;0}\int d^4x  \Phi^{0}\Phi^{0}\Phi^{0}, \qquad \lambda_1=\lambda_2=\lambda_3 =0,
 \end{split}
 \end{equation}
 where $\bar{\mathbb{P}}$ was given in (\ref{3aug2}), while $\mathbb{P}$ is its complex conjugate. Integration by parts needs to be used to see that these have the following symmetry properties under permutations of field labels
 \begin{equation}
 \label{17jul13}
 \begin{split}
 \bar{\mathbb{P}}_{ij} &\equiv \bar\partial_i\beta_j-\bar\partial_j \beta_i=- \bar{\mathbb{P}}_{ji}, \qquad 
  {\mathbb{P}}_{ij} \equiv \partial_i\beta_j-\partial_j \beta_i=- {\mathbb{P}}_{ji},\\
   \bar{\mathbb{P}}&\equiv \bar{\mathbb{P}}_{12}= \bar{\mathbb{P}}_{23}=\bar{\mathbb{P}}_{31}, \qquad 
     {\mathbb{P}}\equiv {\mathbb{P}}_{12}= {\mathbb{P}}_{23}={\mathbb{P}}_{31}.
  \end{split}
 \end{equation}

  Let us briefly pause and make a couple of comments on the result (\ref{17jul12}). 
  Firstly, unless the total helicity in the vertex is zero, for any triplet of helicities there is a single consistent vertex. Instead, when the total helicity is zero,  only the scalar self-interaction vertex is consistent. Secondly, vertices manifestly split into holomorphic and antiholomorphic ones. Namely, when the total helicity is positive, the vertex features positive powers of $\bar{\mathbb{P}}$ and does not depend on $\mathbb{P}$. Analogously, negative helicity vertices depend on $\mathbb{P}$ and not on $\bar{\mathbb{P}}$. Thirdly, at this stage of the analysis the coupling constants $C^{\lambda_1;\lambda_2;\lambda_3}$ are arbitrary. Finally, not all vertices in (\ref{17jul12}) are non-trivial. Namely, due to Bose symmetry and antisymmetry of $\bar{\mathbb{P}}_{ij}$ and ${\mathbb{P}}_{ij}$, vertices with the total helicity odd and involving two identical fields are vanishing. Usually, one goes a bit further by assuming that the coupling constants $C^{\lambda_1;\lambda_2;\lambda_3}$ are symmetric in helicities. This then entails that all vertices with the total helicity odd are vanishing. We will also stick to this assumption.
  
  For fields, which are valued in the adjoint representation of an internal Lie algebra the coupling constants in (\ref{17jul12}) need to be replaced as
\begin{equation}
\label{19jul2}
C^{\lambda_1;\lambda_2;\lambda_3} \to C^{\lambda_1;\lambda_2;\lambda_3} f^{a_1a_2a_3},
\end{equation}
where $f$ are the associated structure constants\footnote{For theories with internal symmetries the free action needs to be dressed with the color indices in an obvious way.}.
We will again assume that the complete coupling constants are symmetric in field labels. Considering that $f^{a_1a_2a_3}$ is totally antisymmetric, we conclude that only vertices with the total helicity being odd are non-trivial. 
Similarly, for the case of  bi-adjoint internal symmetry
each coupling constant should be replaced as 
\begin{equation}
\label{19jul13}
C^{\lambda_1;\lambda_2;\lambda_3} \to C^{\lambda_1;\lambda_2;\lambda_3} f^{a_1a_2a_3} \tilde f^{b_1b_2b_3}.
\end{equation}
Due to the fact that structure constants are totally antisymmetric, vertices are non-trivial for total helicity being even.

\subsection{Second order consistency conditions}

 Constraints imposed by Lorentz invariance at the next order lead to equations, which involve cubic vertices quadratically and quartic vertices linearly.  
  In \cite{Metsaev:1991mt,Metsaev:1991nb} it was observed that these constraints feature two closed subsectors: a sector that involves only holomorphic cubic vertices quadratically and the analogous sector with antiholomorphic vertices only. This observation, in particular, allows one  to fix coupling constants $C^{\lambda_1;\lambda_2;\lambda_3}$ in higher-spin theories \cite{Metsaev:1991mt,Metsaev:1991nb}. 
  Another crucial property of the consistency conditions at this order is that by setting one of the sectors of cubic vertices to zero -- either holomorphic or antiholomorphic ones -- one ends up with a consistent theory. In other words, a cubic theory with vertices of one holomorphicity and  the coupling constants properly fixed turns out to be identically Lorentz invariant and, thus, does not require completion with higher-order interactions. This latter observation can be naturally derived from the analysis of \cite{Metsaev:1991mt,Metsaev:1991nb}, though, it was stated explicitly only in  \cite{Ponomarev:2016lrm}. This peculiar structure of the light-cone deformation procedure is what allows one to construct chiral higher-spin theories avoiding issues with non-locality.

  We will now briefly review how the second order consistency conditions for chiral theories look like. We will follow \cite{Ponomarev:2016lrm}, where these were simplified and a systematic way of solving them was given. As in the case of no internal symmetry and in the cases with internal symmetries of different types the analysis goes differently, these will be considered separately.
  
  \subsubsection{No internal symmetry}
  \label{noin}
  
  The second order consistency conditions for chiral theories involve three exchange-like contributions built out of cubic vertices constructed at the previous step. As a result, these feature momenta of four fields $\Phi^{\lambda_i}$ each associated with one external line of the exchange diagram. It turns out that after the appropriate rearrangements are made, momenta only appear in the following combinations
\begin{equation}
\label{16jul1}
\begin{split}
2A&\equiv \bar{\mathbb{P}}_{12}+\bar{\mathbb{P}}_{34}= -   \bar{\mathbb{P}}_{14}+ \bar{\mathbb{P}}_{23},\\
2B&\equiv \bar{\mathbb{P}}_{13}-\bar{\mathbb{P}}_{24}=    \bar{\mathbb{P}}_{34}- \bar{\mathbb{P}}_{12},\\
2C&\equiv \bar{\mathbb{P}}_{14}+\bar{\mathbb{P}}_{23}= -   \bar{\mathbb{P}}_{13}- \bar{\mathbb{P}}_{24}.
\end{split}
\end{equation}
Equations in the right column of (\ref{16jul1}) are only valid once momentum conservation is taken into account.

The consistency condition that needs to be satisfied acquires the form
\begin{equation}
\label{16jul2}
\begin{split}
\sum_{\omega}'\Big[ &\big((\lambda_1-\lambda_2+\lambda_3-\lambda_4)A+(\lambda_1-\lambda_2-\lambda_3+\lambda_4)B - 2\omega C\big)\\
&\qquad \qquad C^{\lambda_1,\lambda_2,\omega}C^{\lambda_3,\lambda_4,-\omega}(A-B)^{\lambda_1+\lambda_2+\omega-1} (A+B)^{\lambda_3+\lambda_4-\omega-1}\\
+&\big((-\lambda_1+\lambda_2+\lambda_3-\lambda_4)B+(-\lambda_1-\lambda_2+\lambda_3+\lambda_4)C + 2\omega A\big)\\
&\qquad\qquad C^{\lambda_1,\lambda_3,\omega}C^{\lambda_2,\lambda_4,-\omega}(B-C)^{\lambda_1+\lambda_3+\omega-1} (-B-C)^{\lambda_2+\lambda_4-\omega-1}\\
+&\big((-\lambda_1-\lambda_2+\lambda_3-\lambda_4)A+(\lambda_1+\lambda_2-\lambda_3-\lambda_4)C - 2\omega B\big)\\
&\qquad\qquad C^{\lambda_1,\lambda_4,\omega}C^{\lambda_2,\lambda_3,-\omega}(C-A)^{\lambda_1+\lambda_4+\omega-1} (A+C)^{\lambda_2+\lambda_3-\omega-1}\Big]=0.
\end{split}
\end{equation}
Here we use prime over the summation sign to denote that the sum goes only over odd powers of brackets such as $(A+B)$ and $(A-B)$ in the second line. This restriction of the sum to odd terms is a consequence of the fact that vertices with the total helicity being odd in the given case are vanishing, see discussion below (\ref{17jul13}).

Although  equation (\ref{16jul2}) may appear formidable it can be solved systematically as follows. We start by setting  $C=0$, which leads us to the equation of the schematic form
\begin{equation}
\label{16jul3}
(A-\kappa B)f(A,B)=k_{\Lambda-1}A^{\Lambda-1}+k_{\Lambda-2}A^{\Lambda-2}B+ k_1 A B^{\Lambda-2}+k_0 B^{\Lambda-1},
\end{equation}
where 
\begin{equation}
\label{16jul4}
\begin{split}
f(A,B)&\equiv \sum_{\omega}'C^{\lambda_1,\lambda_2,\omega}C^{\lambda_3,\lambda_4,-\omega}(A-B)^{\lambda_1+\lambda_2+\omega-1} (A+B)^{\lambda_3+\lambda_4-\omega-1},\\
\Lambda &\equiv \lambda_1+\lambda_2+\lambda_3+\lambda_4,
\end{split}
\end{equation}
while $k_i$ and $\kappa$ are some $\lambda$-dependent coefficients. Their exact form will not be needed.

Next, considering that only vertices with the total helicity even are non-vanishing, one has
\begin{equation}
\label{16jul5}
f(A,B)= - f(B,A).
\end{equation}
It is relatively straightforward to show that together with (\ref{16jul3}) this leads to
\begin{equation}
\label{16jul6}
f(A,B)= k_{\Lambda-1}(A^{\Lambda-2}-B^{\Lambda-2}).
\end{equation}
It is now easy to see that 
\begin{equation}
\label{16jul7}
C^{\lambda_1,\lambda_2,\lambda_3} = g \frac{l^{\lambda_1+\lambda_2+\lambda_3-1}}{(\lambda_1+\lambda_2+\lambda_3-1)!}
\end{equation}
solves (\ref{16jul4}), (\ref{16jul6}). To this end one just needs to use the binomial expansion formula as well as to keep in mind that only odd power of $(A-B)$ and $(A+B)$ contribute to the sum. Then one checks that (\ref{16jul7}) does solve the original consistency condition (\ref{16jul2}).
We have thus reproduced the coupling constants of the chiral higher-spin theory (\ref{4aug1}).

Another solution to (\ref{16jul2}) is given by 
\begin{equation}
\label{19jul1}
C^{\lambda_1,\lambda_2,\lambda_3} = g \delta^{\lambda_1+\lambda_2+\lambda_3}_2
\end{equation}
and it corresponds to the Poisson limit of the chiral higher-spin theory, (\ref{4aug2}). As we mentioned, a theory (\ref{19jul1}) admits various truncations with self-dual gravity among them. Otherwise, (\ref{16jul7}) and (\ref{19jul1}) provide a general solution to (\ref{16jul2}) for spinning fields with $f(A,B)\ne 0$. Solutions with $f(A,B)=0$ will be briefly discussed in the section devoted to the case of bi-adjoint symmetry. 

Finally, let us remark that 
\begin{equation}
\label{4augv1}
f(A,B) + f(B,C)+f(C,A)=0,
\end{equation}
which holds as a consequence of (\ref{16jul6}), is the Jacobi identity for the structure constants defined in (\ref{4aug4}). Thus, the method of solving (\ref{16jul2}) that we presented above, explains why Lorentz invariance for the theories we considered entails that these also satisfy the BCJ relations in the form discussed in section \ref{BCJsd}.

\subsubsection{Adjoint internal symmetry}

Once fields are valued in the adjoint representation coupling constants need to be replaced as in (\ref{19jul2}).
The consistency conditions get deformed accordingly. To be more precise, (\ref{16jul2}) will receive extra factors of 
\begin{equation}
\label{19jul3}
f^{a_1a_2e}f^{a_3a_4}{}_e, \qquad f^{a_1a_3e}f^{a_2a_4}{}_e \qquad \text{and} \qquad f^{a_1a_4e}f^{a_2a_3}{}_e
\end{equation}
to the second, the fourth and the sixth lines accordingly.
Next, we can use the Jacobi identity 
\begin{equation}
\label{19jul4}
f^{a_1a_2e}f^{a_3a_4}{}_e - f^{a_1a_3e}f^{a_2a_4}{}_e + f^{a_1a_4e}f^{a_2a_3}{}_e=0
\end{equation}
to eliminate the second term in favour of the first and the third one. By equating to zero coefficients in front of the two remaining combinations of $f$ we arrive to two equivalent equations. 
Let us pick the following one
\begin{equation}
\label{19jul5}
\begin{split}
\sum_{\omega}''&\Big[ \big((\lambda_1-\lambda_2+\lambda_3-\lambda_4)A+(\lambda_1-\lambda_2-\lambda_3+\lambda_4)B - 2\omega C\big)\\
&\qquad\qquad\qquad C^{\lambda_1,\lambda_2,\omega}C^{\lambda_3,\lambda_4,-\omega}(A-B)^{\lambda_1+\lambda_2+\omega-1} (A+B)^{\lambda_3+\lambda_4-\omega-1}\\
+&\big((-\lambda_1+\lambda_2+\lambda_3-\lambda_4)B+(-\lambda_1-\lambda_2+\lambda_3+\lambda_4)C + 2\omega A\big)\\
&\qquad\qquad\qquad C^{\lambda_1,\lambda_3,\omega}C^{\lambda_2,\lambda_4,-\omega}(B-C)^{\lambda_1+\lambda_3+\omega-1} (-B-C)^{\lambda_2+\lambda_4-\omega-1}\Big]=0,
\end{split}
\end{equation}
where the sum goes over even powers of $(A+B)$ and $(A-B)$ only, which we denote by a double prime over the summation sign.

It can be solved similarly to (\ref{16jul2}). By setting $C=0$ we find an equation of the form
\begin{equation}
\label{19jul6}
(A-\kappa B)f(A,B) = k_1 AB^{\Lambda-2}+k_0B^{\Lambda-1}.
\end{equation}
We then require that $(A-\kappa B)$ is a divisor of the right-hand side, which leads to
\begin{equation}
\label{19jul7}
f(A,B)=k_0 B^{\Lambda-2}.
\end{equation}
Considering that vertices with the total helicity even vanish, we find that $f(A,B)=f(B,A)$. This is only compatible with (\ref{19jul7}) for $\Lambda=2$. This, in turn, entails that 
\begin{equation}
\label{19jul8}
C^{\lambda_1;\lambda_2;\lambda_2}=g \delta^{\lambda_1+\lambda_2+\lambda_3}_1.
\end{equation}
We have, thus, reproduced the colored version of chiral higher-spin theories (\ref{4aug3}).
As in the case of no color symmetry, (\ref{19jul7}) with $\Lambda=2$ entails the Jacobi identity for the structure constants defined as in (\ref{4aug5}).

\subsubsection{Bi-adjoint internal symmetry}

In a similar way we will consider the case of bi-adjoint internal symmetry, which was not studied before in this language.
This, in particular, will allow us to see why the solution that corresponds to the bi-adjoint scalar does not follow the general pattern of section \ref{BCJsd}, despite being associated with a Lorentz invariant theory.

Substituting (\ref{19jul13}) into (\ref{16jul2}) we should then use the Jacobi identity for both Lie algebras and then collect contributions in front of independent contractions of the Lie algebra structure constants, then setting them to zero. As a result, we find 
\begin{equation}
\label{19jul14}
\begin{split}
\sum_{\omega}'&\Big[ \big((\lambda_1-\lambda_2+\lambda_3-\lambda_4)A+(\lambda_1-\lambda_2-\lambda_3+\lambda_4)B - 2\omega C\big)\\
&\qquad\qquad\qquad C^{\lambda_1,\lambda_2,\omega}C^{\lambda_3,\lambda_4,-\omega}(A-B)^{\lambda_1+\lambda_2+\omega-1} (A+B)^{\lambda_3+\lambda_4-\omega-1}\Big]=0
\end{split}
\end{equation}
and two other equations, which can be obtained by the crossing symmetry. We can again simplify the analysis by considering $C=0$,
which  entails
\begin{equation}
\label{19jul15}
\big((\lambda_1-\lambda_2+\lambda_3-\lambda_4)A+(\lambda_1-\lambda_2-\lambda_3+\lambda_4)B \big)f(A,B)=0.
\end{equation}

Let us first consider solutions for which the prefactor of $f(A,B)$ in (\ref{19jul15}) vanishes. This leads to
\begin{equation}
\label{19jul16}
\lambda_1=\lambda_2, \qquad \lambda_3=\lambda_4.
\end{equation}
Together with the equations related by crossing symmetry, this implies that all $\lambda_i$ are equal, $\lambda_i=\lambda$.
For $\lambda=0$ we, thus, obtain the bi-adjoint scalar theory. 

In the opposite case, $\lambda\ne 0$, we need to check whether (\ref{19jul14}) is satisfied. It is, indeed, the case. This happens due to the fact that for 
\begin{equation}
\label{19jul17}
C^{\lambda,\lambda,\lambda}=g\delta^{\lambda_1}_\lambda \delta^{\lambda_2}_\lambda \delta^{\lambda_3}_\lambda,
\end{equation}
either $C^{\lambda_1,\lambda_2,\omega}$ or $C^{\lambda_3,\lambda_4,-\omega}$ equals zero.
Solutions of this type do exist both for the no internal symmetry case and for the case with the adjoint representations of internal symmetry.

Clearly, there are many other solutions of this type. In general, when none of the vertices in the theory can be connected by a propagator, then (\ref{19jul14}) is trivially satisfied. Theories of this type can be equivalently characterised by $f(A,B)=0$. In particular, for a theory, which has non-trivial vertices only when all helicities in the vertex are positive one has $f(A,B)=0$. These theories have Abelian global symmetries and higher number of derivatives. Besides that, since $C^{1,-1,\lambda}$ and $C^{2,-2,\lambda}$ both have to be vanishing, 
fields in these theories cannot interact minimally neither with Yang-Mills fields nor with gravity. All these properties make them less attractive and  we will not discuss these in what follows.

\section{Copies without internal symmetry}
\label{sec:4}

In this section we will start constructing theories of the type suggested in section \ref{strategy}. As parent theories we will use the chiral higher-spin theory and its Poisson limit, self-dual Yang-Mills theory and self-dual gravity. 
We will first explain how the light-cone deformation procedure applies to this case. Next, we will proceed to the solutions of the associated second order consistency conditions imposed by Lorentz invariance. We will study theories without internal symmetry first. The case of adjoint internal symmetry will be studied in the next section.

\subsection{First order consistency conditions}

One can immediately write the free action for a theory with spectrum (\ref{16jul12})  as
(\ref{17jul10})
  \begin{equation}
 \label{17jul11}
 S_2= \sum_{\nu,\mu,\rho,\dots}\frac{1}{2}\int d^4 x  \Phi^{\lambda,\mu,\rho,\dots} \Box \Phi^{-\lambda,-\mu,-\rho,\dots}.
 \end{equation}
As we mentioned in the previous section, the results of the analysis of the first order consistency conditions do not rely on the spectrum. Instead, cubic vertices are fixed by  helicities of the fields involved and all these come with arbitrary coupling constants. Focusing on the anti-holomorphic sector we find
 \begin{equation}
 \label{17jul14}
 \begin{split}
 S_3 &= C^{\lambda_1,\mu_1,\rho_1,\dots ;\lambda_2,\mu_2\rho_2,\dots;\lambda_3,\mu_3,\rho_3,\dots}\\
 & \int d^4 x \frac{\bar{\mathbb{P}}^{\lambda_1+\lambda_2+\lambda_3+\mu_1+\mu_2+\mu_3+\rho_1+\rho_2+\rho_3+\dots}}{\beta_1^{\lambda_1+\mu_1+\rho_1+\dots}\beta_2^{\lambda_2+\mu_2+\rho_2+\dots}\beta_3^{\lambda_3+\mu_3+\rho_3+\dots}} \Phi^{\lambda_1,\mu_1,\rho_1,\dots}\Phi^{\lambda_2,\rho_2,\mu_2,\dots}\Phi^{\lambda_3,\rho_3,\mu_3,\dots}.
  \end{split}
 \end{equation}
This result is applicable when the total helicity is positive
 \begin{equation}
 \label{17jul15}
 \lambda_1+\lambda_2+\lambda_3+\mu_1+\mu_2+\mu_3+\rho_1+\rho_2+\rho_3+\dots > 0.
 \end{equation}
Similarly, one can write out holomorphic vertices. These will not be needed for our construction.

\subsection{Second order consistency conditions}

As discussed in the previous section, for theories with vertices of one holomorphicity one only needs to solve the consistency condition (\ref{16jul2}) on the coupling constants, which should be properly extended to account for a new spectrum. 
Below we will solve them in different cases.

\subsubsection{Copies of chiral higher-spin theories}

We start by analysing the double copy of chiral higher-spin theories. The consistency condition (\ref{16jul2}) gets replaced with
\begin{equation}
\label{24jul1}
\begin{split}
\sum_{\omega,\nu}'&\Big[ \big((\lambda_1+\mu_1-\lambda_2-\mu_2+\lambda_3 +\mu_3-\lambda_4-\mu_4)A\\
&\qquad +(\lambda_1+\mu_1-\lambda_2-\mu_2-\lambda_3-\mu_3+\lambda_4+\mu_4)B 
- 2(\omega+\nu) C\big)\\
& C^{\lambda_1,\mu_1;\lambda_2,\mu_2;\omega,\nu}C^{\lambda_3,\mu_3;\lambda_4,\mu_4;-\omega,-\nu}\\
&\qquad \qquad (A-B)^{\lambda_1+\mu_1+\lambda_2+\mu_2+\omega-\nu-1} (A+B)^{\lambda_3+\omega_3+\lambda_4+\omega_4-\omega-\nu-1}\\
&\qquad\qquad\qquad\qquad\qquad \qquad\qquad+(2\leftrightarrow 3) + (2\to4,3\to 2,4\to 3) \Big] =0.
\end{split}
\end{equation}
As before, prime over the summation sign indicates that the sum is taken over odd powers of $(A-B)$ and $(A+B)$ only. By repeating the arguments from section \ref{noin}, one finds that (\ref{24jul1}) entails
\begin{equation}
\label{17jul1}
\begin{split}
\sum_{\omega,\nu}' & C^{\lambda_1,\mu_1;\lambda_2,\mu_2;\omega,\nu}  C^{\lambda_3,\mu_3;\lambda_4,\mu_4;-\omega,-\nu}\\
&\qquad  (A-B)^{\lambda_1+\mu_1+\lambda_2+\mu_2+\omega+\nu-1} (A+B)^{\lambda_3+\mu_3+\lambda_4+\mu_4-\omega-\nu-1}\\
&\qquad \qquad \qquad \qquad \qquad \qquad \qquad \qquad   =k (A^{\Lambda+M-2}-B^{\Lambda+M-2}),
\end{split}
\end{equation}
where 
\begin{equation}
\label{20jul1}
M\equiv \mu_1+\mu_2+\mu_3+\mu_4
\end{equation}
and $\Lambda$ was defined in (\ref{16jul4}).

It would be interesting to solve (\ref{17jul1}) in complete generality. In the present paper, however, we use a restricted setup motivated by the double-copy procedure. Namely, we require that the dependence of vertices on $\lambda_i$ and $\mu_i$ factorises, (\ref{4aug6}). Kinematic dependence of vertices on $\lambda_i$ and $\mu_i$ factorises automatically, see (\ref{17jul14}). Thus, we only need to impose
\begin{equation}
\label{5aug1}
C^{\lambda_1,\mu_1;\lambda_2,\mu_2;\lambda_3,\mu_3}=C_{\cal L}^{\lambda_1;\lambda_2;\lambda_3}C_{\cal M}^{\mu_1;\mu_2;\mu_3}.
\end{equation}
Focusing on solutions of this type, we find
\begin{equation}
\label{17jul2}
C^{\lambda_1,\mu_1;\lambda_2,\mu_2;\lambda_3,\mu_3} = g \frac{l^{\sum \lambda-1-a}}{(\sum \lambda-1-a)!}
 \frac{m^{\sum\mu+a}}{(\sum\mu+a)!},
\end{equation}
where $a$ is an arbitrary integer number and
\begin{equation}
\label{5aug2}
\sum \lambda \equiv \lambda_1+\lambda_2+\lambda_3, \qquad \sum\mu \equiv \mu_1+\mu_2+\mu_3.
\end{equation}
To see that (\ref{17jul2}) does solve (\ref{17jul1}) one can notice that if the sum is taken over all $\omega$ and $\nu$, then it factorises and can be easily evaluated with the binomial expansion formula. To take into account that the sum actually goes only over odd powers of $(A-B)$ and $(A+B)$ we may just project the answer into its part, which is odd under permutations of $A$ and $B$. Of course, one also needs to check that (\ref{17jul2}) solves the original equation (\ref{24jul1}). This is done explicitly in appendix \ref{app:a}.

A few comments are now in order. Firstly, formula (\ref{17jul2}) features an arbitrary integer parameter $a$. Its presence can be easily explained as follows. Unlike in the single copy of the chiral higher-spin theory, in the double copy labels $\lambda$ and $\mu$ alone do not bare any physical meaning. Indeed, only helicity $h=\lambda+\mu$ controls how the fields transform under Lorentz transformations and, thus, contributes to the consistency conditions, while $\lambda$ or $\mu$ separately just account for the spectrum degeneration. Due to that, solutions to (\ref{24jul1}) should go over to solutions under shifts of $\lambda$ and $\mu$ that keep their sum invariant. An arbitrary parameter $a$ in (\ref{17jul2}) is just a manifestation of this symmetry. Alternatively, this means that $a$ in (\ref{17jul2}) does not have any physical meaning and can be set to any convenient integer number, say, $a=0$.

Secondly, we would like to point out the following subtlety. For  solution (\ref{17jul2}) to work, it is important that the sum in (\ref{17jul1}) goes over all terms with powers of $(A-B)$ and $(A+B)$ being odd. This means that all vertices with the total helicity $\sum\lambda+\sum\mu$ even should be non-trivial. 
These, in particular, include vertices for which both $\sum\lambda$ and $\sum\mu$ are odd. 
We remind the reader, that vertices of this type in single copy chiral higher-spin theories vanish. Thus, the double-copy theory obtained above does not simply reduce to the multiplication of the coupling constants and vertices in individual theories, but it also requires to supplement the product with the additional vertices for which both $\sum\lambda$ and $\sum\mu$ are odd\footnote{In chiral higher-spin theories with matrix-valued fields \cite{Skvortsov:2018jea,Skvortsov:2020wtf}  all vertices are non-vanishing. This happens due to the fact that in contrast to the color Lie algebra structure constants, traces of products of matrices are not totally antisymmetric.
Thus, higher-spin theories with matrix-valued fields involve an extension of more familiar color symmetries of a Lie algebra type, which, moreover, relies on their interplay with kinematic symmetries. It seems natural to expect that the presence of this unified structure already in the single copy theory can streamline the double copy procedure and, in particular, will allow one to avoid the necessity to extend single copies with additional vertices. }.

Finally, we note that the double copy chiral higher-spin theory (\ref{17jul2}) is a new non-trivial theory in the sense that it cannot be reduced to the sum of single-copy theories by linear field redefinitions. This is shown in appendix \ref{app:b}.

In a similar manner one can consider multiple products of chiral higher-spin theories. Consistency condition (\ref{24jul1}) changes in an obvious manner. Solving it in the same way as before, we generate a family of multiple copy chiral higher-spin theories. 
For example, the couple constants of the triple-copy chiral higher-spin theory  read 
\begin{equation}
\label{17jul3}
C^{\lambda_1,\mu_1,\rho_1;\lambda_2,\mu_2,\rho_2;\lambda_3,\mu_3,\rho_3} = g \frac{l^{\sum\lambda-1}}{(\sum\lambda-1)!}
 \frac{m^{\sum\mu}}{(\sum\mu)!} \frac{r^{\sum\rho}}{(\sum\rho)!},
\end{equation}
where we fixed the freedom of redefinitions of $\lambda$, $\mu$ and $\rho$ in a convenient way. More generally, introducing a generalised version of the first line in (\ref{4aug4}) as
\begin{equation}
\label{5aug4}
\begin{split}
n^x_{CHS}\equiv \frac{l^{\sum\lambda-x}}{(\sum\lambda-x)!}\frac{\bar{\mathbb{P}}^{\sum\lambda-1}}{\beta_1^{\lambda_1-x}\beta_2^{\lambda_2-x}\beta_3^{\lambda_3+x}},
\end{split}
\end{equation}
 vertices for the $p$-fold copy chiral higher-spin theories can be schematically given as
\begin{equation}
\label{5aug5}
\text{CHS}^p: \qquad V_3 = n \cdot n^1_{CHS} \big( n^0_{CHS}\big)^{p-1}.
\end{equation}

\subsubsection{Copies of chiral and Poisson higher-spin theories}

One may wonder whether chiral higher-spin multiple copies admit limits analogous to the one that leads to  (\ref{19jul1}) for the chiral higher-spin theory.
Considering that for multiple-copy theories one has multiple coupling constants that can be sent to zero, -- such as $l$, $m$ and $r$ in (\ref{17jul3}) -- one has a number of options how these limits can be performed. Below, instead of performing such limits, we will study the second order consistency conditions separately, assuming that cubic  vertices have factors analogous to (\ref{19jul1}).

To start, let us consider the case of the Poisson chiral higher-spin double copy. To this end, we solve the consistency condition (\ref{17jul1}) with the additional requirement $\Lambda+M-2=2$, which mimics an analogous condition for the single copy theory.
 It is not hard to see that the associated solution reads
\begin{equation}
\label{1aug1}
C^{\lambda_1,\mu_1;\lambda_2,\mu_2;\lambda_3,\mu_3} = g\delta_2^{\sum\lambda}\delta_0^{\sum\mu},
\end{equation}
where we fixed the shift symmetry of $\mu$ and $\lambda$ in a convenient way.
This result can be straightforwardly extended to the case of multiple copies. For example, the coupling constants for the triple copy theory read
\begin{equation}
\label{1aug2}
C^{\lambda_1,\mu_1,\rho_1;\lambda_2,\mu_2,\rho_2;\lambda_3,\mu_3,\rho_3} = g\delta_2^{\sum\lambda}\delta_0^{\sum\mu}  \delta_0^{\sum\rho}.
\end{equation}

In a similar way we consider theories, for which the Poisson limit is taken only for some factors. In particular, by solving the consistency conditions one
finds the coupling constants that correspond to the product  of two chiral higher-spin theories, with one of them in the Poisson limit,
\begin{equation}
\label{1aug3}
C^{\lambda_1,\mu_1;\lambda_2,\mu_2;\lambda_3,\mu_3} = g\frac{l^{\sum\lambda-1}}{(\sum\lambda-1)!}\delta_0^{\sum\mu}.
\end{equation}
Its generalisation to the case of higher powers of each factor can be easily found as well. For example, for quintuple copy of the chiral higher-spin theory with two factors in the Poisson limit, we find the following coupling constants
\begin{equation}
\label{1aug4}
C^{\lambda_1,\mu_1,\rho_1,\sigma_1,\chi_1;\dots}=g\frac{l^{\sum\lambda-1}}{(\sum\lambda-1)!}
 \frac{m^{\sum\mu}}{(\sum\mu)!} \frac{r^{\sum\rho}}{(\sum\rho)!}
\delta_0^{\sum\sigma}
\delta_0^{\sum\chi}.
\end{equation}

 Results of this section can be summarised with the aid of the deformed Poisson factor
\begin{equation}
\label{6aug1}
n^x_{PCHS}\equiv \delta_x^{\sum\lambda}\frac{\bar{\mathbb{P}}^{x}}{\beta_1^{\lambda_1-x}\beta_2^{\lambda_2-x}\beta_3^{\lambda_3+x}}.
\end{equation}
Then, vertices for the $p$-fold copy Poisson chiral higher-spin theories can be schematically given as
\begin{equation}
\label{6aug2}
\text{PCHS}^p: \qquad V_3 = n \cdot n^1_{PCHS} \big( n^0_{PCHS}\big)^{p-1}.
\end{equation}
For products of powers of chiral higher-spin theories with powers of the Poisson factors we have
\begin{equation}
\label{6aug3}
\text{CHS}^p\times\text{PCHS}^q: \qquad V_3 = n \cdot n^1_{CHS} \big( n^0_{CHS}\big)^{p-1} \big( n^0_{PCHS}\big)^{q}.
\end{equation}

As mentioned before, the only chiral higher-spin theory for which the original double copy procedure does suggest a natural double copy is the theory with internal symmetries (\ref{4aug3}). In this case, by factoring out the internal structure constants and by squaring the kinematic part of the vertex, we obtain $\delta_1^{\sum\lambda}\delta_1^{\sum\mu}$ as the coupling constants. We can then relabel helicities as discussed above to find that this theory is equivalent to the one given by (\ref{1aug1}). Thus, the latter theory can be equivalently regarded as the double copy of (\ref{4aug3}). We would also like to emphasise that (\ref{1aug1}) features a closed subsector, consisting of fields with $\mu_i=0$. This subsector is nothing else but the Poisson chiral higher-spin theory (\ref{4aug2}) and its role is completely analogous to that of the gravitational subsector of the theory, which, in addition to gravity features the dilaton and the Kalb-Ramon field (\ref{3aug12}). Summarising this discussion, we find that for the only case to which the original double copy is naturally applicable, it works exactly as expected.

\subsubsection{Products of  chiral higher-spin theories, self-dual Yang-Mills theory and self-dual gravity}

In a similar manner we  consider theories, which besides the chiral higher-spin theory involve self-dual Yang-Mills theory and self-dual gravity as factors.

To start, we consider a product of the chiral higher-spin theory and self-dual Yang-Mills theory. In the present section we are dealing with colorless theories, so only contributions from the kinematic part of the self-dual Yang-Mills vertex, (\ref{3aug13}), will be included into the product of vertices. Our recipe  from section \ref{strategy} implies that we should consider a theory with the spectrum of fields  $\Phi^{\lambda,\mu}$, where $\lambda\in \mathbb{Z}$, while $\mu = \pm 1$. Keeping in mind that the only non-trivial vertex in the Yang-Mills theory has $\sum\mu=1$, we find that (\ref{24jul1}) reduces to
\begin{equation}
\label{17jul4}
\begin{split}
\sum_{\omega}C_{\cal L}^{\lambda_1;\lambda_2;\omega}C_{\cal L}^{\lambda_3;\lambda_4,;-\omega}
(A-B)^{\lambda_1+\lambda_2+\omega} (A+B)^{\lambda_3+\lambda_4-\omega}\\
=k (A^{\Lambda}-B^{\Lambda}),
\end{split}
\end{equation}
where we used 
\begin{equation}
\label{17jul8}
\mu_1+\mu_2+\nu = 1, \qquad \mu_3+\mu_4-\nu=1
\end{equation}
to evaluate the $\nu$ summation.
Equation (\ref{17jul4}) has an obvious solution, which leads to
\begin{equation}
\label{17jul5}
C^{\lambda_1,\mu_1;\lambda_2,\mu_2;\lambda_3,\mu_3} = g \frac{l^{\sum\lambda}}{(\sum\lambda)!}
(\delta^{\mu_1}_1\delta^{\mu_2}_1\delta^{\mu_3}_{-1}+\delta^{\mu_1}_1\delta^{\mu_2}_{-1}\delta^{\mu_3}_{1}+\delta^{\mu_1}_{-1}\delta^{\mu_2}_1\delta^{\mu_3}_{1}).
\end{equation}

It is instructive to rewrite the result (\ref{17jul5}) in a slightly different way. To this end, let us introduce new notations for fields
\begin{equation}
\label{6aug4}
\Psi^{\lambda+1,+}\equiv \Phi^{\lambda,1}, \qquad \Psi^{\lambda-1,-}\equiv \Phi^{\lambda,-1}.
\end{equation}
In other words, as the first field label we use the total helicity of the field, while the second label refers to the helicity of the Yang-Mills factor. Then, (\ref{17jul5}) corresponds to the following vertex
\begin{equation}
\label{6aug5}
 V_3=3g\sum_{h_i \in \mathbb{Z}}\frac{l^{h_1+h_2+h_3-1}}{(h_1+h_2+h_3-1)!}\int d^4 x \frac{\bar{\mathbb{P}}^{h_1+h_2+h_3}}{\beta_1^{h_1}\beta_2^{h_2}\beta_3^{h_3}} \Psi^{h_1,+}\Psi^{h_2,+}\Psi^{h_3,-}.
\end{equation}
For each triplet of helicities it has the same structure constants as the chiral higher-spin theory we started from (\ref{4aug1}).
However, theory (\ref{6aug5}) has twice as many fields as the chiral higher-spin theory. This doubled spectrum has two sectors: the one that involves all $\Psi^{h,+}$ and the other one, that involves all $\Psi^{h,-}$.
  From (\ref{6aug5}) one can conclude that each vertex is bilinear in fields from the $+$ sector and linear in fields of the $-$ sector. This implies that the $+$ sector is closed under evolution, while fields from the $-$ sector propagate on the background provided by the $+$ sector fields.
 The existence of two field sectors, which enter asymmetrically into interactions is a characteristic feature of self-dual Yang-Mills theory and self-dual gravity\footnote{
 Here we refer to (\ref{3aug1}) and (\ref{3aug5}) as self-dual Yang-Mills theory and self-dual gravity respectively, which feature fields of opposite helicities. The presence of both signs of helicity is necessary for the action to be Lorentz invariant. If one relaxes the requirement of Lorentz invariance or the requirement of a theory to have an action, then self-dual Yang-Mills theory and self-dual gravity can be consistently truncated to the sector involving only fields of positive helicities.
 }, while it is absent in the chiral higher-spin theory in the original form (\ref{16jul7}). Thus,  (\ref{6aug5}) may be regarded as a simple extension of the chiral higher-spin theory which makes it structurally more similar to self-dual lower-spin theories.

In a similar way one can construct theories with higher powers of higher-spin contributions. For example, the product of two higher-spin theories and self-dual Yang-Mills theory leads to the coupling constants
\begin{equation}
\label{17jul6}
\begin{split}
C^{\lambda_1,\mu_1,\rho_1;\lambda_2,\mu_2,\rho_2;\lambda_3,\mu_3,\rho_3}  = g \frac{l^{\sum\lambda}}{(\sum\lambda)!}
 \frac{m^{\sum\mu}}{(\sum\mu)!}
 (\delta^{\rho_1}_1\delta^{\rho_2}_1\delta^{\rho_3}_{-1}+\delta^{\rho_1}_1\delta^{\rho_2}_{-1}\delta^{\rho_3}_{1}+\delta^{\rho_1}_{-1}\delta^{\rho_2}_1\delta^{\rho_3}_{1}).
\end{split}
\end{equation}
This theory is related to the double copy of the chiral higher-spin theory (\ref{17jul2}) in the same way as (\ref{6aug5}) is related to the chiral higher-spin theory.

Alternatively, on can consider theories that involve  higher powers of the Yang-Mills contributions. For example, the chiral higher-spin theory times self-dual Yang-Mills theory squared leads to
\begin{equation}
\label{17jul7}
\begin{split}
& C^{\lambda_1,\mu_1,\rho_1;\lambda_2,\mu_2,\rho_2;\lambda_3,\mu_3,\rho_3}  = g \frac{l^{\sum\lambda+1}}{(\sum\lambda+1)!}\\
& \qquad \qquad (\delta^{\mu_1}_1\delta^{\mu_2}_1\delta^{\mu_3}_{-1}+\delta^{\mu_1}_1\delta^{\mu_2}_{-1}\delta^{\mu_3}_{1}+\delta^{\mu_1}_{-1}\delta^{\mu_2}_1\delta^{\mu_3}_{1})
 (\delta^{\rho_1}_1\delta^{\rho_2}_1\delta^{\rho_3}_{-1}+\delta^{\rho_1}_1\delta^{\rho_2}_{-1}\delta^{\rho_3}_{1}+\delta^{\rho_1}_{-1}\delta^{\rho_2}_1\delta^{\rho_3}_{1}).
\end{split}
\end{equation}
This theory is yet another extension of the chiral higher-spin theory, which has four times as many fields as the original theory.
Theories with multiple chiral higher-spin theory and self-dual Yang-Mills factors can be constructed analogously. 

It is worth noting that if the higher-spin factors are absent, self-dual Yang-Mills theories can be consistently copied at most twice. Thus, one cannot construct consistent copies of self-dual Yang-Mills theory beyond self-dual gravities. At the technical level of our consistency conditions this can be explained as follows. Once three Yang-Mills factors are present in the theory, this already brings the overall factor of $(A-B)^2(A+B)^2$ to $f(A,B)$.
At the same time, Lorentz invariance requires that $f(A,B)$ has the form $A^\Lambda-B^\Lambda$ for some $\Lambda$. Since $(A-B)^2(A+B)^2$ is not a divisor of $A^\Lambda-B^\Lambda$ for any $\Lambda$, Lorentz invariance cannot be achieved, unless the remaining sums in $f(A,B)$ feature negative powers of $(A-B)$ and $(A+B)$.
These negative powers can only contribute from the higher-spin factor, once the argument of the factorial in the denominator is properly adjusted.
Let us emphasise that the presence of negative powers of $(A-B)$ and $(A+B)$ at these intermediate steps does not imply that the theory is non-local: though, powers of $\bar{\mathbb{P}}$ associated with the higher-spin factor are negative, the total powers of $\bar{\mathbb{P}}$ in the vertex, once the Yang-Mills contributions are included, are positive.

In a similar way one can treat products of chiral higher-spin theories with self-dual gravity. For example, the product of the single copy of the chiral higher-spin theory and the single copy of gravity leads to the coupling constants
\begin{equation}
\label{17jul9}
C^{\lambda_1,\mu_1;\lambda_2,\mu_2;\lambda_3,\mu_3} = g \frac{l^{\sum\lambda+1}}{(\sum\lambda+1)!}
(\delta^{\mu_1}_2\delta^{\mu_2}_2\delta^{\mu_3}_{-2}+\delta^{\mu_1}_2\delta^{\mu_2}_{-2}\delta^{\mu_3}_{2}+\delta^{\mu_1}_{-2}\delta^{\mu_2}_2\delta^{\mu_3}_{2}).
\end{equation}
By simple field redefinitions, this theory can be brought to the form (\ref{6aug5}). Overall, considering that self-dual gravity is itself the double copy of self-dual Yang-Mills theory, a separate analysis of products of chiral higher-spin theories and self-dual gravity is not needed.

The cubic vertex for the product of any power of chiral higher-spin theories with any power of self-dual Yang-Mills theory can be  schematically written as
\begin{equation}
\label{6aug6}
\text{CHS}^p\times\text{YM}^q: \qquad V_3 = n \cdot n^{1-q}_{CHS} \big( n^0_{CHS}\big)^{p-1} \big( n'\big)^{q}.
\end{equation}
We can also add powers of the Poisson chiral higher-spin theories as a factor, which leads to
\begin{equation}
\label{26sep1}
\text{CHS}^p\times \text{PCHS}^r\times\text{YM}^q : \qquad V_3 = n \cdot n^{1-q}_{CHS} \big( n^0_{CHS}\big)^{p-1} \big( n^0_{PCHS}\big)^{r}\big( n'\big)^{q}.
\end{equation}
This formula summarises our results for theories without internal symmetries.

\section{Copies with internal symmetry in the adjoint representation}
\label{sec:5}

In this section we will consider theories with fields valued in the adjoint representation of the internal symmetry algebra. All vertices then get supplemented with the internal Lie algebra structure constant as a factor. Extending the associated consistency condition (\ref{19jul5}) to the case of the tensor product spectrum with two  factors, we find
\begin{equation}
\label{19jul9}
\begin{split}
\sum_{\omega,\nu}''& C^{\lambda_1,\mu_1;\lambda_2,\mu_2;\omega,\nu}C^{\lambda_3,\mu_3;\lambda_4,\mu_4;-\omega,-\nu}\\
&(A-B)^{\lambda_1+\mu_1+\lambda_2+\mu_2+\omega+\nu-1} (A+B)^{\lambda_3+\mu_3+\lambda_4+\mu_4-\omega-\nu-1}\\
& \qquad \qquad\qquad\qquad\qquad\qquad\qquad \qquad\qquad \qquad=k A^{\Lambda+M-2}.
\end{split}
\end{equation}
The summation above goes over even powers of $(A-B)$ and $(A+B)$, which is consistent with (\ref{19jul9}) only for  $\Lambda+M=2$.

Let us focus on the case of the square of chiral higher-spin theories. Assuming factorisation of vertices, we find
\begin{equation}
\label{19jul10}
C^{\lambda_1,\mu_1;\lambda_2,\mu_2;\lambda_3,\mu_3}=g \delta^{\sum\lambda}_1
 \delta^{\sum\mu}_0.
\end{equation}
In a similar way, we can construct multiple copies of higher-spin theories with adjoint internal symmetry. For example, for the triple copy we find
\begin{equation}
\label{19jul11}
C^{\lambda_1,\mu_1,\rho_1;\lambda_2,\mu_2,\rho_2;\lambda_3,\mu_3,\rho_3}=g \delta^{\sum\lambda}_1
 \delta^{\sum\mu}_0 \delta^{\sum\rho}_0.
\end{equation}

Next, we consider theories with  the Yang-Mills factors involved. For example,  one and two Yang-Mills factors can be added to  the colored chiral higher-spin theory as follows
\begin{equation}
\label{19jul12}
\begin{split}
C^{\lambda_1,\mu_1;\lambda_2,\mu_2;\lambda_3,\mu_3}&=g \delta^{\sum\lambda}_0
 (\delta^{\mu_1}_1\delta^{\mu_2}_1\delta^{\mu_3}_{-1}+\delta^{\mu_1}_1\delta^{\mu_2}_{-1}\delta^{\mu_3}_{1}+\delta^{\mu_1}_{-1}\delta^{\mu_2}_1\delta^{\mu_3}_{1}),\\
C^{\lambda_1,\mu_1,\rho_1;\lambda_2,\mu_2,\rho_2;\lambda_3,\mu_3,\rho_3}&=g \delta^{\sum\lambda}_{-1}
 (\delta^{\mu_1}_1\delta^{\mu_2}_1\delta^{\mu_3}_{-1}+\delta^{\mu_1}_1\delta^{\mu_2}_{-1}\delta^{\mu_3}_{1}+\delta^{\mu_1}_{-1}\delta^{\mu_2}_1\delta^{\mu_3}_{1})\\
 & \qquad\qquad (\delta^{\rho_1}_1\delta^{\rho_2}_1\delta^{\rho_3}_{-1}+\delta^{\rho_1}_1\delta^{\rho_2}_{-1}\delta^{\rho_3}_{1}+\delta^{\rho_1}_{-1}\delta^{\rho_2}_1\delta^{\rho_3}_{1}).
\end{split}
\end{equation}
As in the case on no internal symmetry, these vertices are exactly the same as in higher-spin theory we started from except that the spectrum of the theory becomes doubled, quadrupled, etc.
Gravitational factors can be added to the vertex in a similar fashion.
These results can be schematically summarised as
\begin{equation}
\label{6aug7}
\text{CCHS}^p\times\text{YM}^q: \qquad V_3 = n \cdot n^{1-q}_{CCHS} \big( n^0_{PHS}\big)^{p-1} \big( n'\big)^{q},
\end{equation}
where we introduced 
\begin{equation}
\label{6aug8}
n^x_{CCHS}\equiv f^{abc}\delta_x^{\sum\lambda}\frac{\bar{\mathbb{P}}^{x}}{\beta_1^{\lambda_1-x}\beta_2^{\lambda_2-x}\beta_3^{\lambda_3+x}}.
\end{equation}
Let us note that in $\text{CCHS}^p$ only the kinematic part of the vertex gets raised to power $p$, while  the color structure constants enter only once. Due to this, this factor can be  alternatively regarded as $\text{CCHS} \times \text{PCHS}^{p-1} $.

\section{Symmetry algebras}
\label{sec:syma}

In the previous sections we constructed families of chiral higher-spin theories. For the case of no internal symmetry our results are summarised by 
(\ref{26sep1}), which gives vertices for the product of any powers of the chiral higher-spin theory, its Poisson limit and self-dual Yang-Mills theory. For the case of fields valued in the adjoint representation of some internal Lie algebra our results are summarised by (\ref{6aug7}). It defines a cubic vertex for a theory, which can be regarded as the product of any powers of the colored chiral higher-spin theory and self-dual Yang-Mills theory. Expressions (\ref{26sep1}) and (\ref{6aug7}) have a factorised structure, thus, these can be arguably considered as manifestations of the generalised double copy. 

Another important feature of (\ref{26sep1}) and (\ref{6aug7}) is that these vertices satisfy a version of the BCJ pattern reviewed in section \ref{BCJsd}.
This follows from the very way we arrived to these results. Indeed,  equations such as (\ref{17jul1}) and (\ref{19jul9}) are just the Jacobi identities for the structure constants  entering the BCJ relations in the form  (\ref{4aug5}). Below we will illustrate this phenomenon explicitly as well as obtain Lie algebras associated with some of the theories considered above. We will find that the generalised double copy works even more naturally when considered at the level of these Lie algebras. 

Let us focus on the example of the chiral higher-spin double copy (\ref{17jul2}) with $a=0$. To extract a structure constant from a vertex, we first need to ''raise an index'' of the first field. This is done with a two-form, which is extracted from the kinetic term and simply amounts to changing the signs of momenta and helicity labels of the first field. By dividing the result by the kinematics part of the self-dual Yang-Mills vertex, we find
\begin{equation}
\label{26sep2}
\begin{split}
&[E_2^{\lambda_2;\mu_2}(x),E_3^{\lambda_3;\mu_3}(x)]\Big|_{\lambda_1;\mu_1}
\\
&=\frac{\bar{\mathbb{P}}^{-\check{\lambda}_1+\check{\lambda}_2+\check{\lambda}_3}}{(-\check{\lambda}_1+\check{\lambda}_2+\check{\lambda}_3)!(-\beta_1)^{-\check{\lambda}_1}\beta_2^{\check{\lambda}_2}\beta_3^{\check{\lambda}_3}} 
\frac{\bar{\mathbb{P}}^{-{\mu}_1+{\mu}_2+{\mu}_3}}{(-{\mu}_1+{\mu}_2+{\mu}_3)!(-\beta_1)^{-{\mu}_1}\beta_2^{{\mu}_2}\beta_3^{{\mu}_3}} 
E_2^{\lambda_2;\mu_2}(x)E_3^{\lambda_3;\mu_3}(x),
\end{split}
\end{equation}
where 
\begin{equation}
\label{26sep3}
\check{\lambda}_i \equiv \lambda_i -1, \qquad \beta_1 = -\beta_2-\beta_3.
\end{equation}
In (\ref{26sep2}) $E_i$ are understood as Lie algebra generators and the right-hand side of (\ref{26sep2}) gives the contribution of their commutator to the $(\lambda_1;\mu_1)$ sector. 

It is convenient to rescale the generators as
\begin{equation}
\label{26sep4}
\tilde{E}^{\lambda_i,\mu_i}(x)\equiv \beta_i^{-\check{\lambda}_i -\mu_i}{E}^{\lambda_i,\mu_i}(x)
\end{equation}
and then combine them into a generating function
\begin{equation}
\label{26sep5}
\tilde{E}(x,z,w) \equiv \sum_{\lambda,\mu = -\infty}^{\infty}\tilde{E}^{\lambda_i,\mu_i}(x) z^{\check{\lambda}}w^{\mu}.
\end{equation}
In these terms the commutator simply reads as
\begin{equation}
\label{26sep6}
[\tilde{E}_2(x,z,w),\tilde{E}_3(x,z,w)] = \sinh \left( \left[\frac{l}{z}+\frac{m}{w} \right](\bar{\partial}_2\partial_3^+ -\partial_2^+ \bar{\partial}_3)\right)\tilde{E}_2(x,z,w)\tilde{E}_3(x,z,w),
\end{equation}
where we took into account that only odd powers of $\bar{\mathbb{P}}$ contribute.
The bracket defined in (\ref{26sep6}) clearly satisfies the Jacobi identity. Indeed, this is just the Moyal bracket with $\frac{l}{z}+\frac{m}{w}$ as the non-commutativity parameter\footnote{Rephrasing of (\ref{17jul1}) as the requirement that the bracket $[\tilde{E}_2(x,z,w),\tilde{E}_3(x,z,w)]$ satisfies the Jacobi identity gives an efficient way of solving it. In particular, as well-known, to ensure the Jacobi identity the dependence of the bracket on ${\mathbb{P}}$ should be constrained to that of the Moyal or the Poisson bracket. In the case of the Moyal bracket, the non-commutativity parameter can be any function of the total homogeneity degree $-1$ in $z$ and $w$. The latter requirement follows from Lorentz invariance at the leading order in the coupling constant. Eventually, the bracket gets fixed up to an arbitrary function of $w/z$. This freedom can be potentially removed if one takes into account field redefinitions mentioned in appendix \ref{app:b}, thus, leaving us with a unique algebra (\ref{26sep6}). 
 It would be interesting to explore this issue more thoroughly in future.}. 

It is instructive to compare this result with a Lie algebra associated with the chiral higher-spin theory. In that case one has 
\begin{equation}
\label{26sep7}
[\tilde{E}_2(x,z),\tilde{E}_3(x,z)] = \sinh \left( \frac{l}{z}(\bar{\partial}_2\partial_3^+ -\partial_2^+ \bar{\partial}_3)\right)\tilde{E}_2(x,z)\tilde{E}_3(x,z).
\end{equation}
Thus, double copying of chiral higher-spin theories at the Lie algebra level amounts to the extension of generators with an additional complex variable as well as to the replacement of the non-commutativity parameter as
\begin{equation}
\label{26sep8}
 \frac{l}{z} \to \frac{l}{z}+\frac{m}{w}.
\end{equation}
The multiple copy of chiral higher-spin theories works analogously\footnote{ Relation between Lie algebras (\ref{26sep7}) and (\ref{26sep6}) has some features of the tensor product.
 The possibility to construct chiral higher-spin theories with algebras of the tensor product type was mentioned in \cite{Sharapov:2022wpz}. It would be interesting to clarify whether our results are related. }.

In a similar way we consider the product of the chiral higher-spin theory and its Poisson limit. As a result, we obtain a bracket
\begin{equation}
\label{26sep9}
[\tilde{E}_2(x,z,w),\tilde{E}_3(x,z,w)] = \sinh \left( \frac{l}{z} (\bar{\partial}_2\partial_3^+ -\partial_2^+ \bar{\partial}_3)\right)\tilde{E}_2(x,z,w)\tilde{E}_3(x,z,w).
\end{equation}
Thus, an additional Poisson factor leads to the loop extension of the original algebra.
In turn, by adding a self-dual Yang-Mills factor to the chiral higher-spin theory we find 
\begin{equation}
\label{26sep10}
[\tilde{E}_2(x,z,\theta),\tilde{E}_3(x,z,\theta)] = \sinh \left( \frac{l}{z} (\bar{\partial}_2\partial_3^+ -\partial_2^+ \bar{\partial}_3)\right)\tilde{E}_2(x,z,\theta)\tilde{E}_3(x,z,\theta),
\end{equation}
where $\theta$ is a Grassmann variable $\theta^2=0$. Multiple loop extensions and extensions with multiple Grassmann variables work in an obvious way. Such extensions are also possible for theories with fields in the adjoint representation of internal symmetry.

\section{Conclusion}
\label{sec:6}

In the present paper we studied the double copy  and its generalisations to chiral higher-spin theories. We suggested a procedure that allows one to multiply any powers of self-dual theories. At the level of the little group spectrum, this procedure simply reduces to the tensor multiplication of the spectra of the theories involved. To mimic how the standard double copy works we also required that cubic vertices -- which are the only interactions for chiral theories in the light-cone gauge -- factorise into, possibly, deformed vertices of the parent theories. These vertices were then  fixed by  Lorentz invariance.

As a result, we constructed new higher-spin theories, which include powers of the chiral higher-spin theory and its Poisson limit as well as powers of chiral higher-spin theories with fields in the adjoint representation of the internal symmetry algebra. We also constructed products of these theories with powers of self-dual Yang-Mills theory and self-dual gravity. The only chiral higher-spin theory to which the standard double copy procedure is naturally applicable is the theory with internal symmetries acting in the adjoint representation. We found that in this case our procedure gives the result which is identical to what can be expected from the straightforward extrapolation of the lower-spin double copy. This example solidifies our procedure as a generalisation of the lower-spin double copy. At the same time, our procedure extends well beyond this specific case and, thus, allows us to substantially extend  the existing duality web of double-copy-related theories.

An important common feature of the resulting theories is that their vertices are always given by the product of the kinematic part of the Yang-Mills vertex and some Lie algebra structure constant, which we prefer to regard as a version of the BCJ relations. The associated Jacobi identity arises at intermediate steps in the process of solving the consistency conditions imposed by Lorentz invariance. An alternative argument proving that such a structure of vertices follows from Lorentz invariance was given in \cite{Ponomarev:2017nrr}\footnote{It is quite remarkable, that in this analysis the BCJ relations are derived from Lorentz invariance alone without any references to underlying string theory constructions, as in the standard proofs of color-kinematics duality, see e. g. \cite{Bjerrum-Bohr:2010pnr}. 
We should emphasise, however, that this discussion is only applicable to the self-dual sector, which is much simpler than complete parity-invariant theories. It would be certainly interesting to see whether our results can be extended away from the self-dual sector.}, where it was also shown that this implies that all chiral theories can be rewritten as a self-dual Yang-Mills theory with a, possibly, space-time symmetry instead of internal symmetry. As argued there -- see also \cite{Chacon:2020fmr,Monteiro:2022lwm,Monteiro:2022xwq} -- all theories possessing such a form of the BCJ structure are integrable and have tree-level amplitudes vanishing. One can also anticipate that loop amplitudes in these theories are vanishing after the appropriate regularisation of the sum over helicities is carried out, see \cite{Skvortsov:2018jea,Skvortsov:2020wtf}.

In the course of our analysis we constructed a series of chiral higher-spin theories, thus, substantially extending  a not very numerous  set of higher-spin theories known. The qualitative feature of theories that  we obtained is that their spectra are richer than the typical higher-spin spectrum  containing a field of each spin only once.
Such an extension is beneficial for the following reason.
 The standard higher-spin spectrum is known to have difficulties with higher-spin symmetry breaking. 
 This can already be seen from a superficial counting of degrees of freedom carried by massless and massive fields. At the same time, higher-spin symmetry is not observed, while massive particles are, indeed, present in nature. This means that higher-spin theories with the standard spectrum cannot describe an unbroken phase of a realistic theory of nature, unless, their spectrum is extended.  This problem motivated developments of higher-spin theories with larger spectra \cite{Vasiliev:2012tv,Vasiliev:2018zer}.
An alternative approach to this issue consists in carrying out various versions of dimensional reduction, which immediately result in the rich spectrum of massive fields \cite{Skvortsov:2020pnk,Didenko:2023txr}. In this context, higher-spin theories that we constructed  do have extended spectra and, thus, might be able to accommodate various symmetry breaking patterns.
It would be interesting to explore this issue in future as well as a potential connection to the unbroken phase of string theory\footnote{Certainly, an important drawback of theories that we are dealing with here is that these are of self-dual type. At the same time, the problem of symmetry breaking primarily deals with the spectrum, so one may analyse it, at least, at the qualitative level before the parity-invariant completion of higher-spin theories is known.}.

Finally, let us emphasise, once again, that our analysis was motivated by the double-copy procedure and, thus, we considered only theories of a special type. Namely, we considered only theories for which both the spectrum and the vertices factorise. It would be interesting to consider theories with more general spectra as well as to find the general solution to the consistency conditions imposed by Lorentz invariance. By now, such an analysis was only done systematically for theories without spectrum degeneracy. There is an intriguing possibility, that a systematic study of theories with degenerate spectra  will show that the landscape of massless higher-spin theories is far richer than currently known.

\acknowledgments

We would like to acknowledge discussions with P. Pichini, which contributed to our interest to the problem of the higher-spin double copy. We also thank E. Skvortsov for useful comments on the draft.

\appendix

\section{Conventions}
\label{app:0}

In this section we review our conventions.
We use the light-cone coordinates
\begin{equation}
\label{3aug3}
\begin{split}
x^+& \equiv \frac{1}{\sqrt{2}}(x^3+x^0), \qquad  x^- \equiv \frac{1}{\sqrt{2}}(x^3-x^0),\\
x &\equiv \frac{1}{\sqrt{2}}(x^1-ix^2),\qquad \bar{x} \equiv \frac{1}{\sqrt{2}}(x^1+ix^2)
\end{split}
\end{equation}
for which the metric reads
\begin{equation}
\label{3aug4}
ds^2 = 2 dx^+ dx^- + 2 dx d\bar{x}.
\end{equation}

We impose the light-cone gauge as it is usually done in the higher-spin literature. Namely, fields $\Phi^{+s}$ and $\Phi^{-s}$, carrying helicities $+s$ and $-s$ respectively, are connected to the higher-spin potential $\varphi$ as
\begin{equation}
\label{3sep1}
\Phi^{+s}=\varphi^{x(s)}, \qquad \Phi^{-s}=\varphi^{\bar{x}(s)}.
\end{equation}
Here e. g. $\varphi^{x(s)}$ is a component of the totally symmetric rank-s tensor $\varphi$ with all indices taking value $x$.
All the non-vanishing components of $\varphi$ are given in the following way
\begin{equation}
\label{3sep2}
\varphi^{-(n)I(s-n)}= \left(-\frac{\partial_J}{\partial^+} \right)\varphi^{J(n)I(s-n)},
\end{equation}
where $I$ and $J$ run over $x$ and $\bar x$. For example, for the Maxwell field for which only the helicity $+1$ component is non-vanishing, we have 
\begin{equation}
\label{3sep3}
\varphi^+=0, \qquad \varphi^x=0, \qquad \varphi^{\bar x} = \Phi^{+1}, \qquad \varphi^- = -\frac{\partial_x}{\partial^+}\Phi^{+1}.
\end{equation}

Along with these conventions, alternative ones are often used,  see e. g. \cite{Monteiro:2011pc}. In these alternative conventions the coordinates are chosen as
\begin{equation}
\label{3sep4}
\begin{split}
u &\equiv x^0-x^3 = -\sqrt{2} x^-, \qquad  v \equiv x^0+x^3=\sqrt{2}x^+,\\
w &\equiv x^1+ix^2=\sqrt{2}\bar{x},\qquad \bar{w} \equiv x^1-ix^2=\sqrt{2}x.
\end{split}
\end{equation}
Moreover, the light-cone gauge is imposed with different normalisations. For the example we considered above, one has
\begin{equation}
\label{3sep5}
\varphi_u=0, \qquad \varphi_w=0, \qquad \varphi_v=-\frac{1}{4}\partial_w\phi, \qquad \varphi_{\bar w}=-\frac{1}{4}\partial_u\phi.
\end{equation}
Employing (\ref{3aug3}), (\ref{3sep3}), (\ref{3sep5}) it is not hard to establish the connection between the two field choices
\begin{equation}
\label{3sep6}
\Phi^{+1}=-\frac{1}{2\sqrt{2}}\partial_u\phi =\frac{1}{4}\partial^+ \phi.
\end{equation}
For arbitrary positive helicities this connection reads
\begin{equation}
\label{3sep7}
\Phi^{+s}=-\frac{1}{4}(\sqrt{2})^s(\partial_u)^s\phi =-\frac{1}{4}(-1)^s(\partial^+)^s \phi.
\end{equation}
For negative helicities the formula is analogous. Note that powers of $\partial^+$ on the right-hand side of (\ref{3sep7}) cancel exactly the powers of $\beta$ entering cubic vertices (\ref{17jul12}). This is consistent with the expressions for lower-spin vertices written in terns of $\phi$ appearing in  \cite{Monteiro:2011pc}.

Finally, let us note that unlike \cite{Monteiro:2011pc} we use the mostly plus signature.

\section{Explicit check of the double copy solution}
\label{app:a}

In this appendix we check that vertices (\ref{17jul2}) satisfy the complete consistency condition, not only its consequence (\ref{17jul1}).
To be more precise, the complete consistency condition reads (\ref{24jul1}).
We will check that it is satisfied by 
\begin{equation}
\label{24jul2}
C^{\lambda_1,\mu_1;\lambda_2,\mu_2;\lambda_3,\mu_3} = \frac{1}{(\lambda_1+\lambda_2+\lambda_3-1)!}\frac{1}{(\mu_1+\mu_2+\mu_3)!}.
\end{equation}

To start, we will first sum the necessary terms over all $\omega$ and $\nu$. Then the sum factorises and it is easy to evaluate that
\begin{equation}
\label{24jul3}
\begin{split}
&\sum_{\omega,\nu} \big((\lambda_1+\mu_1-\lambda_2-\mu_2+\lambda_3 +\mu_3-\lambda_4-\mu_4)A\\
&\qquad \qquad +(\lambda_1+\mu_1-\lambda_2-\mu_2-\lambda_3-\mu_3+\lambda_4+\mu_4)B )\\
&\qquad \frac{(A-B)^{\lambda_1+\lambda_2+\omega-1}}{(\lambda_1+\lambda_2+\omega-1)!}\frac{(A+B)^{\lambda_3+\lambda_4-\omega-1}}{(\lambda_3+\lambda_4-\omega-1)!}
\frac{(A-B)^{\mu_1+\mu_2+\nu}}{(\mu_1+\mu_2+\nu)!}\frac{(A+B)^{\mu_1+\mu_2-\nu}}{(\mu_1+\mu_2-\nu)!}\\
&\qquad =\big((\lambda_1+\mu_1-\lambda_2-\mu_2+\lambda_3 +\mu_3-\lambda_4-\mu_4)A\\
&\qquad\qquad +(\lambda_1+\mu_1-\lambda_2-\mu_2-\lambda_3-\mu_3+\lambda_4+\mu_4)B )
\frac{(2A)^{\Lambda+M-2}}{(\Lambda-2)!M!}
\end{split}
\end{equation}
and
\begin{equation}
\label{24jul4}
\begin{split}
&-2C\sum_{\omega,\nu} (\omega+\nu)
\frac{(A-B)^{\lambda_1+\lambda_2+\omega-1}}{(\lambda_1+\lambda_2+\omega-1)!}\frac{(A+B)^{\lambda_3+\lambda_4-\omega-1}}{(\lambda_3+\lambda_4-\omega-1)!}\\
&\qquad \qquad\qquad\qquad\qquad\qquad \frac{(A-B)^{\mu_1+\mu_2+\nu}}{(\mu_1+\mu_2+\nu)!}\frac{(A+B)^{\mu_1+\mu_2-\nu}}{(\mu_1+\mu_2-\nu)!}\\
&\qquad =
-2C \frac{(2A)^M}{M!}\left(-B \frac{(2A)^{\Lambda-3}}{(\Lambda-3)!}-\frac{\lambda_1+\lambda_2-\lambda_3-\lambda_4}{2}\frac{(2A)^{\Lambda-2}}{(\Lambda -2)!} \right)\\
&\qquad\qquad-2C \frac{(2A)^{\Lambda-2}}{(\Lambda-2)!}\left(-B \frac{(2A)^{M-1}}{(M-1)!}-\frac{\mu_1+\mu_2-\mu_3-\mu_4}{2}\frac{(2A)^{M}}{M!} \right).
\end{split}
\end{equation}

Now we need to take into account that the sum goes over odd powers of $(A-B)$ and $(A+B)$ only. This can be easily implemented by antisymmetrising the summation results in $A \leftrightarrow B$. Note that in (\ref{24jul3}) the bracket in the prefactor should not be antisymmetrised. Indeed, it does not come from the sum, instead, it was originally present as an $\omega$- and $\nu$-independent prefactor.

By carrying out this antisymmetrisation and then adding the cross-channel terms, we find that the result adds up to zero. Thus, the consistency condition (\ref{24jul1}) for the coupling constants (\ref{24jul2}) is, indeed, satisfied.

\section{Non-triviality of the double copy}
\label{app:b}

Theories like (\ref{17jul2}) feature infinite sets of fields of the same helicity, which we can label as
\begin{equation}
\label{24jul5}
\Phi^{\lambda,\mu} =\Phi^{h+d,h-d}, \qquad d\in \mathbb{Z}.
\end{equation}
Consider linear field redefinitions within the set of fields of the same helicity
\begin{equation}
\label{24jul6}
\Phi^{h+d,h-d}= \sum_i M^{(h)}_{d,i}\tilde{\Phi}^{h,i}.
\end{equation}
These redefinitions leave the helicity invariant, thus, they leave the consistency conditions imposed by Lorentz invariance unchanged. This, in turn, means that along with the solution (\ref{17jul2}) for the coupling constants, we also have an infinite set of solutions, which are generated by redefinitions (\ref{24jul6}).

Considering this symmetry, it seems natural to ask whether (\ref{17jul2}) is, indeed, a theory in which all fields $\Phi^{\lambda,\mu}$ are non-trivially mixed by interactions or, instead, it is just an infinite sum of theories, such as the chiral higher-spin theory we started with, the latter property being obscured by a redefinition of the form (\ref{24jul6}). In other words, we would like to check whether one can use field redefinitions to diagonalise the coupling constants
\begin{equation}
\label{24jul7}
\tilde{C}^{h_1,i_1;h_2,i_2;h_3,i_3}=\sum_{d_1,d_2,d_3} {C}^{h_1+d_1,h_1-d_1;h_2+d_2,h_2-d_2;h_3+d_3,h_3-d_3}M^{(h_1)}_{d_1,i_1}M^{(h_2)}_{d_2,i_2}M^{(h_3)}_{d_3,i_3}
\end{equation}
in the sense that
\begin{equation}
\label{24jul8}
\tilde{C}^{h_1,i_1;h_2,i_2;h_3,i_3}=0 \qquad \text{unless} \qquad i_1=i_2=i_3.
\end{equation}

Below we will show that (\ref{24jul8}) cannot be achieved. To do that, we will show that a weaker property
\begin{equation}
\label{24jul9}
\tilde{C}^{h_1,i_1;h_1,i_2;h_3,i_3}=0 \qquad \text{unless} \qquad i_1=i_2
\end{equation}
cannot be satisfied for all $h_3$ and $i_3$. Note that we set first two helicities equal.
This can be rephrased as a statement that 
\begin{equation}
\label{24jul10}
{C}^{h_1+d_1,h_1-d_1;h_1+d_2,h_1-d_2;h_3+d_3,h_3-d_3},
\end{equation}
which transforms with $M^{(h_1)}_{d_1,i_1}$ on the first two sets of indices cannot be simultaneously diagonalised for all $h_3$ and $d_3$.

Mathematically, this problem reduces to the problem of simultaneous diagonalisation of an infinite set of quadratic forms labelled by $h_3$ and $d_3$.
From the principal axis theorem one knows that the basis that diagonalises a quadratic form is given by eigenvectors of this quadratic form regarded as a linear map. Thus, the question whether quadratic forms can be simultaneously diagonalised reduces to the problem of the existence of a basis of simultaneous eigenvectors of the associated maps. In turn, the latter property is equivalent to the requirement that all these maps commute.
In other words, we need to check whether
\begin{equation}
\label{24jul11}
\begin{split}
&\sum_{d'}{C}^{h_1+d_1,h_1-d_1;h_1+d',h_1-d';h_3+d_3,h_3-d_3}{C}^{h_1+d',h_1-d';h_1+d_2,h_1-d_2;h'_3+d'_3,h'_3-d'_3}\\
&-\sum_{d'}{C}^{h_1+d_1,h_1-d_1;h_1+d',h_1-d';h'_3+d'_3,h'_3-d'_3}{C}^{h_1+d',h_1-d';h_1+d_2,h_1-d_2;h_3+d_3,h_3-d_3}=0.
\end{split}
\end{equation}
Here pairs $(h_3,d_3)$ and $(h'_3,d'_3)$ label matrices that we commute, while $d_1$ and $d_2$ are the matrix indices of the commutator that is set to zero. In other words, (\ref{24jul11}) needs to be satisfied for all $d_1$, $d_2$, $d_3$, $d'_3$, $h_1$ and all even $h_3$ and $h'_3$. 

We computed the left-hand side of (\ref{24jul11}) for some particular values of variables and found that it does not vanish. Thus, vertices (\ref{17jul2}) cannot be diagonalised by field redefinitions.

It would be also interesting to check in a similar way whether there are relations between the theories we constructed in the present paper. In particular, it may happen that the chiral higher-spin double copy can be brought to the form of the product of the chiral higher-spin theory and the Poisson chiral higher-spin theory factor when field redefinitions (\ref{24jul6}) are used. 
We leave a systematic analysis of this problem for future research.

\bibliography{hsmulty}
\bibliographystyle{JHEP}

\end{document}